\documentclass[showpacs, pra, aps, twocolumn,10pt]{revtex4}
\usepackage{mathrsfs}
\usepackage{array}
\usepackage{amsmath}
\usepackage{amssymb}
\usepackage{graphicx}
\usepackage{float}
\usepackage{amstext}
\usepackage{bm}
\usepackage{amsfonts}
\usepackage{color}

\DeclareMathOperator{\sech}{sech}

\begin{document}

\title{Optomechanical non-Gaussian quantum steering and remote preparation of large-size motional Sch\"{o}rdinger cat states}
\author{Miaomiao Wei}
\affiliation{Department of Physics, Huazhong Normal University, Wuhan 430079, China}
\author{Huatang Tan}
\email{tht@mail.ccnu.edu.cn}
\affiliation{Department of Physics, Huazhong Normal University, Wuhan 430079, China}
\begin{abstract}
In this paper, we present a scheme for remotely generating large-size motional Schr\"{o}dinger cat states in cavity optomechanical (OM) systems with  non-Gaussian quantum steering of continuous variables. We consider that the output field from the OM cavity undergoes three typical kinds of multiphoton operations: multiphoton subtraction, multiphoton addition, or multiphoton catalysis, followed by homodyne detection. We first demonstrate that these multiphoton operations can lead to non-Gaussian OM quantum steerable correlations, which are unveiled by the subsequent homodyne detection with a Fisher-information-based steering criterion. It is found that the non-Gaussian steering is obviously enhanced with an increasing number $n$ of photons in the multiphoton operations, which, as we show, fails to be revealed with the well-known Reid's steering criterion. It therefore suggests that the Fisher-information-based criterion is more effective for witnessing non-Gaussian quantum steering.
We next show that the strong OM steering enables the remote preparation of large-size Schr\"{o}dinger odd or even cat states of the mechanical oscillator by the homodyne detection. Accordingly, the amplitudes of the cat states also increase significantly with the photon number $n$, particularly in the cases of multiphoton subtraction and addition. Our results reveal the properties of non-Gaussian steering generated by multiphoton operations, and the large cat states of macroscopic mechanical resonators hold promise for fundamental tests in quantum mechanics and practical applications in quantum science.
\end{abstract}
\maketitle

\section{Introduction}

Non-Gaussian quantum states \cite{NGS}, which may exhibit negative Wigner quasiprobability functions that indicate genuine nonclassicality, can offer specific advantages in various applications in quantum science, including quantum computation \cite{qcom}, quantum communication \cite{tp}, and quantum metrology \cite{estimation}. Among the non-Gaussian operations used to generate non-Gaussian states, photon addition \cite{addition1,addition2} and photon subtraction \cite{subtraction1,subtraction2,subtraction3} are particularly notable. Further, multiphoton subtraction (MPS) \cite{multips1,multips2}, multiphoton addition (MPA)  \cite{multipa}, and multiphoton catalysis (MPC) \cite{multipc1,multipc2, multipc3}, as typical non-Gaussian multiphoton operations, have been demonstrated experimentally to effectively achieve non-Gaussian nonclassical states.
The generation of large-size Schr\"{o}dinger cat states always attracts intense research interests, as they not only contribute to understanding the fundamentals of quantum mechanics \cite{catqm1,catqm2,catqm3,catqm4,catqm5,catqm6}, but also they play a crucial role in a number of quantum technologies \cite{catqt1,catqt2,catqt3,catqt4,catqt5,catqt6,catqt7}, e.g., in quantum computation circuits where coherent states are required as logical qubits  \cite{exc}.
On the other hand, the remote preparation of desirable states \cite{rsp1,rsp2,st}, which relies on entanglement resources, is particularly interesting and has garnered significant attention due to its advantages, such as enabling the remote manipulation of quantum states, enhancing security, consuming less classical information, and requiring no Bell state measurement, compared to direct state transmission and quantum teleportation \cite{com1,com2,com3, spe,loc2}.

Quantum steering provides profound insights into directional quantum nonlocality and delineates the contributions of each subsystem, distinguishing it from the symmetric entanglement characteristic of both observers. A salient feature of quantum steering is its capacity for the two parties to verify the distribution of shared entanglement, even when the measurement devices of one observer are deemed unreliable \cite{Wiseman,RMP}. This property underscores the significance of quantum steering from both theoretical and experimental perspectives. Consequently, it plays a pivotal role in various quantum information protocols, including semi-device-independent quantum key distribution \cite{qk}, quantum secret sharing \cite{qss}, and one-way quantum computing \cite{ow}. Recent studies have further demonstrated that quantum steering is also an important resource for remote state preparation \cite{spe}.
For continuous variable systems, theoretical and experimental studies  of quantum steering generation have primarily focused on Gaussian states \cite{XY}.
Characterizing non-Gaussian quantum steering presents challenges, as a key difficulty lies in the complex correlations manifested in higher-order moments of observables \cite{ho1,ho2,ho3,ho4,ho5}. Moreover, the experimental detection of these higher-order moments is also challenging. Recently, the characterization of non-Gaussian steering with quantum fisher information (FI) via a metrological protocal by using homodyne detection has been proposed \cite{complementarity,Homodyne}.

In recent decades, extensive research and development have established cavity OM systems as an excellent platform for investigating nonclassical Gaussian and non-Gaussian states of photons and phonons \cite{COMPP2, COMPP3, COMPP4}. These non-classical quantum states enable the exploration of quantum phenomena on macroscopic scales and also facilitate quantum tasks \cite{ Macro1, Macro2, Macro3, COMPP6, COMPP7}. Generally, achieving non-Gaussian mechanical states relies on the nonlinearity induced with strong single-photon OM coupling comparable to the mechanical frequency --- a challenge that remains difficult to achieve experimentally. Alternately, phonon subtraction or addition via photon detection are considered an efficient approach to realize non-Gaussian mechanical states \cite{Tan,Li,Xiang}. Very recent experiments have successfully generated vibrational non-Gaussian states via phonon subtraction or addition operations on initially prepared mechanical states \cite{mechanical1,mechanical2,mechanical3}. However, in these studies for achieving non-Gaussian mechanical states, the phonon subtraction or addition operations, performed via the OM beam-splitter interaction and photon detection, depends on specific mechanical initial states such as squeezed or coherent states. The detection of a photon heralds the subtraction  (addition) of a phonon from (to) the initial mechanical states through the beam splitter interaction.

In contrast, in this paper we first consider the realization of OM non-Gaussian steering via the MPS, MPA or MPC operations on the output field from the OM cavity, and then we utilize this non-Gaussian steering to remotely generating mechanical large-size Schr\"{o}dinger cat states via homodyne detection. Therefore, our scheme solely depends on the nonlocal quantum correlations, which renders the remote state preparation. This is achieved not via the aforementioned OM beam splitter interaction but two-mode squeezing interaction and the multiphoton operations. Therefore, our scheme  does not require for initial nonclassical mechanical states.
We further reveal that the FI-based criterion is more effective for witnessing non-Gaussian quantum steering than the well-known Reid's steering. We also find that the non-Gaussian steering and the amplitudes of the Schr\"{o}dinger odd or even cat states is obviously enhanced with an increasing number of photons involved in the multiphoton operations.

This paper is arranged as follows.
In Sec. II, the fisher-information steering criterion is reviewed in brief. In Sec. III, the cavity optomechanical system driven by pulse and continuous lasers is introduced. In Sec. IV, the OM non-Gaussian steering via the multiphoton operations on the OM cavity output field is investigated. In Sec.V, the remote generation of large-size mechanical cat states is studied. In Sec. VI, the conclusion is given.

\section{FI-based steering witness via homodyne detection}
\begin{figure}
\centerline{\hspace{0cm}\scalebox{0.3}{\includegraphics{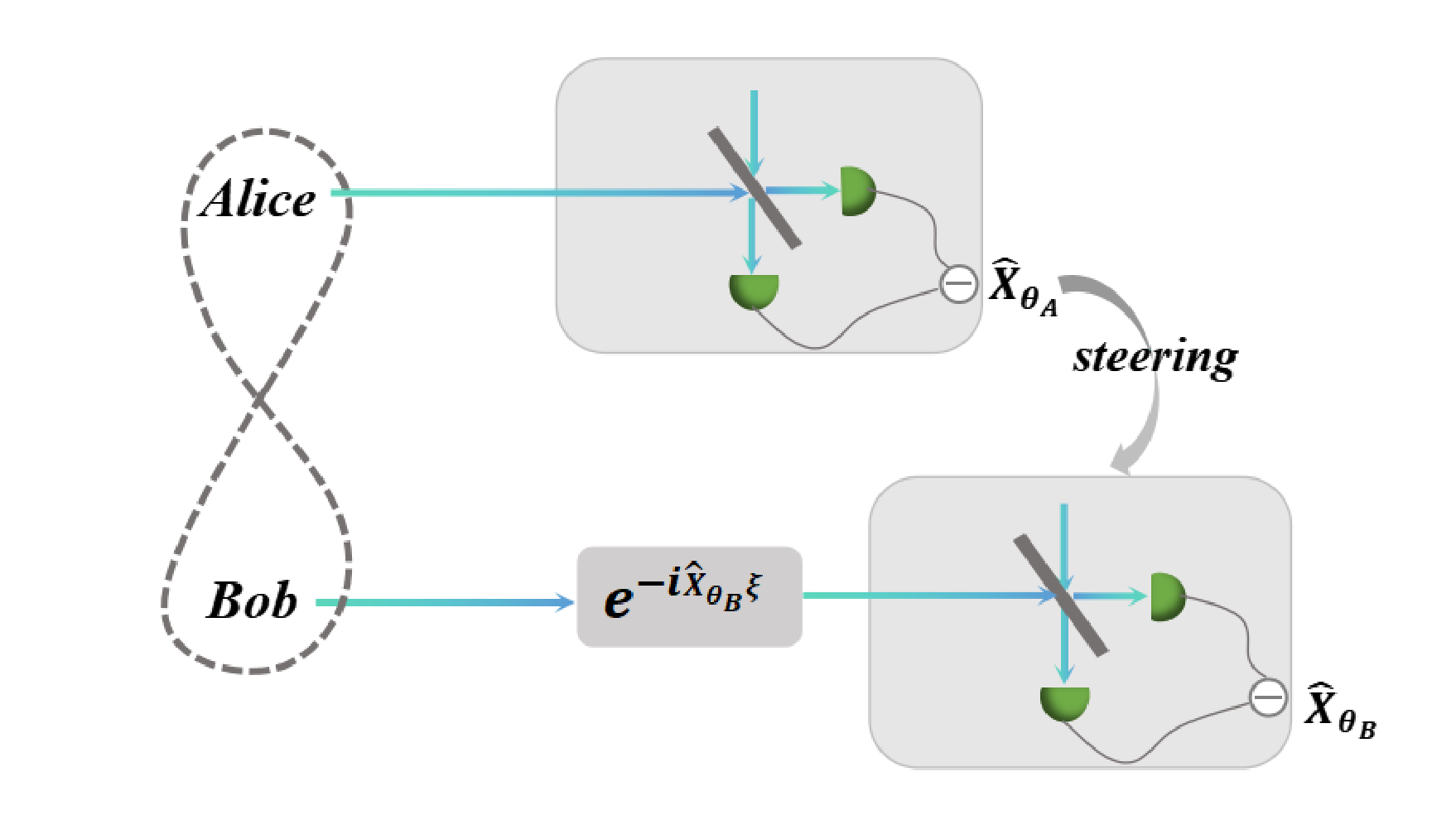}}}
\caption{Metrological-protocol-based witnessing of bipartite quantum steering of continuous variables. Alice performs homodyne
detection on his mode and communicates to Bob the quadrature she chose to measure and its outcome. Based on this information, Bob tunes the local oscillator to choose what's quadrature to measure in order to better estimate the displacement $\xi$ generated by $\hat D(\xi)=e^{-i\hat X_{\theta_B}\xi}$. }
\label{fig1}
\end{figure}
Before we study in detail the properties of the non-Gaussian light-mechanical steering in OM systems, we first briefly review the criterion for non-Gaussian steering with quantum FI, based on quantum metrological protocol and homodyne detection \cite{complementarity,Homodyne}.
In the formulation of the EPR paradox as a metrological task, a local phase shift $\xi$ is generated by the operator $\hat X_{\theta_B}$ on Bob's state, as shown in Fig.\ref{fig1}.
Without any further information than that he can extract from direct measurements in the displaced state $\hat\rho_\xi^B =e^{-i\hat X_{\theta_B}\xi}\hat{\rho}^B e^{i\hat X_{\theta_{\scriptscriptstyle B}}\xi}$, the maximal precision is limited by the quantum Cram\'{e}r-Rao bound
\begin{align}
V(\xi_{est})\geq\frac{1}{MF_Q(\hat{\rho}^B,\hat X_{\theta_B})}.
\label{Rao}
\end{align}
Here, $M$ is the repetition number of measurement and $F_Q(\hat{\rho}^B,\hat X_{\theta_B})$ is the quantum FI which represents the sensitivity of state $\hat\rho_B$ under
small perturbations generated by $\hat X_{\theta_B}$ \cite{metrology1,metrology2,metrology3}.

In the assisted estimation protocol, Alice assists Bob in his estimation protocol by sending him information about her measurement outcomes, and Alice's assistance may improve Bob's estimation precision with the correlations between them. It has been shown in Ref.\cite{complementarity} that local complementarity sets a limit to this improvement that can only be overcome when there exists quantum steerable correlations. The average sensitivity attainable by Bob following assistance by Alice performing homodyne detection on his mode, is upper bounded by the conditional quantum FI
\begin{align}
F_{Q}^{B|A}(\mathcal A,\hat X_{\theta_B})&=\displaystyle\max_{\hat X_{\theta_A}}\int p(x_{\theta_A}\mid\hat X_{\theta_A})\nonumber\\
&~~~~~~~~~~~~\times F_{Q}^{B}(\hat\rho_{x_{\theta_A}\mid\hat X_{\theta_A}}^B,\hat X_{\theta_B})dx_{\theta_A}.
\end{align}
and $\rho_{x_{\theta_A}\mid\hat X_{\theta_A}}^B$ is the reduced state of Bob's mode conditioned on the Alice's measurement outcome.
The assemblage
\begin{align}
\mathcal A(x_{\theta_A},\hat X_{\theta_A})=p(x_{\theta_A}\mid\hat X_{\theta_A})\hat\rho_{x_{\theta_A}\mid\hat X_{\theta_A}}^B,
\end{align}
where $p(x_{\theta_A}\mid\hat X_{\theta_A})$ is the probability distribution for Alice's outcomes $x_{\theta_A}$ for the observable $\hat X_{\theta_A}$.
The confirmation of quantum steering is to show that assemblage (3) cannot be described with a hidden state model \cite{Wiseman}, i.e.,
\begin{align}
\mathcal A(x_{\theta_A},\hat X_{\theta_A})=\sum_{\lambda} p(x_{\theta_A}\mid\hat X_{\theta_A},\lambda)p(\lambda)\hat\sigma_{\lambda}^B.
\end{align}
If the state Bob and Alice share is consistent with the structure of Eq.(4), the following inequality holds \cite{complementarity}:
\begin{align}
F_{Q}^{B|A}(\mathcal A,\hat X_{\theta_B})\leq 4V_{Q}^{B|A}(\mathcal A,\hat X_{\theta_B}),
\label{e6}
\end{align}
where
\begin{align}
V_{Q}^{B|A}(\mathcal A,\hat X_{\theta_B})&=\displaystyle \min_{\hat X_{\theta_A}}\displaystyle\sum_a p(x_{\theta_A}\mid\hat X_{\theta_A})
V(\hat\rho_{x_{\theta_A}\mid\hat X_{\theta_A}}^B,\hat X_{\theta_B}),
\label{cv}
\end{align}
with the variance $V(\hat\rho_{x_{\theta_A}\mid\hat X_{\theta_A}}^B,\hat X_{\theta_B}):=\langle \hat X_{\theta_B}^2\rangle_{\hat\rho_{x_{\theta_A}\mid\hat X_{\theta_A}}^B}-\langle \hat X_{\theta_B}\rangle_{\hat\rho_{x_{\theta_A}\mid\hat X_{\theta_A}}^B}^2$ of the quadrature $\hat X_{\theta_B}$ in the conditional state $\hat\rho_{x_{\theta_A}\mid\hat X_{\theta_A}}^B$.
This inequality can be thought of as a way to witness steering, i.e.,
\begin{align}
S_{\max}^{A\rightarrow B}(\mathcal A)=&\displaystyle\max_{\hat X_{\theta_B}} \Big[F_{Q}^{B|A}(\mathcal A,\hat X_{\theta_B})-4V_{Q}^{B|A}(\mathcal A,\hat X_{\theta_B})\Big]^{+},
\label{e9}
\end{align}
where $[x]^{+}=\max\{0,x\}$.

For realistic homodyne detection on the quadratures $\hat X_{B(A)}$, the quantum FI is lower bounded by the classical counterpart
\begin{align}
F\big[P(x_B\mid\xi)\big]=\int dx_B P(x_B\mid\xi)\big[\partial_{\xi}\log P(x_B\mid\xi)\big]^2,
\end{align}
where $\log P(x_B\mid\xi)$ represents the logarithmic likehood associated with the probability density of measurement outcomes $x_B$, after implementation of the parameters $\xi$.

With the generalized quadrature $\hat X_{\theta_{B}}$ of the Bob's mode denoted by the bosonic operator $\hat o_B$,
\begin{align}
\hat X_{\theta_{B}}=\cos{\theta_{B}}\hat q_{B}+\sin{\theta_{B}}\hat p_{B},
\label{qudt}
\end{align}
where $q_{B}=(\hat o_B+\hat o_B^\dag)/\sqrt{2}$ and $p_{B}=(\hat o_B^\dag-\hat o_B)/\sqrt{2}i$, and similarly for
the generalized quadrature $\hat X_{\theta_{A}}$ of the Alice's mode,
the maximum FI optimized over on Alice's homodyne detection results can be expressed by
\begin{align}
F_{hom}^{B|A}(\mathcal A,\hat X_{\theta_B})=&\displaystyle\max_{\theta_A\in[0,2\pi)}\int P(x_{\theta_A})\nonumber\\
&~~~~~~~~~~~\times F\big[P_{x_{\theta_A}}^B(x_{\theta_B}\mid\xi)\big]dx_{\theta_A}.
\label{e11}
\end{align}
Here $P(x_{\theta_A})$ is the probability distribution along the quadrature measured by Alice, and the conditional probability $P_{x_{\theta_A}}^B(x_{\theta_B}\mid\xi)
=P_{x_{\theta_A}}^B(x_{\theta_B}-\xi)$ dependent on the detection results $x_{\theta_A}$. The conditional variance
\begin{align}
V_{hom}^{B|A}(\mathcal A,{\hat X_{\theta_B}})=&\displaystyle\min_{\theta_A\in[0,2\pi)}\int P(x_{\theta_A}) V(\hat\rho_{x_{\theta_A}}^B,\hat X_{\theta_B})dx_{\theta_A},
\label{e12}
\end{align}
where $V(\hat\rho_{ x_{\theta_A}}^B,{\hat X_{\theta_B}})=\int x_{\theta_B}^2 P_{x_{\theta_A}}^B( x_{\theta_B})d x_{\theta_B}-(\int x_{\theta_B} P_{x_{\theta_A}}^B(x_{\theta_B})dx_{\theta_B})^2$.

With Eqs.(\ref{e11}) and (\ref{e12}), the steering condition (\ref{e9}) with the homodyne detection becomes into
\begin{align}
S_{FI}^{{A\rightarrow B}}(\mathcal A)=\displaystyle\max_{\theta_B\in[0,2\pi)}\Big[F_{hom}^{B|A}(\mathcal A,{\hat X_{\theta_B}})-4V_{hom}^{B|A}(\mathcal A,{\hat X_{\theta_B}})\Big]^{+},
\label{e13}
\end{align}
which is employed in the following to witness non-Gaussian steering.

\begin{figure}
\vspace{0cm}
\centerline{\hspace{0cm}\scalebox{0.35}{\includegraphics{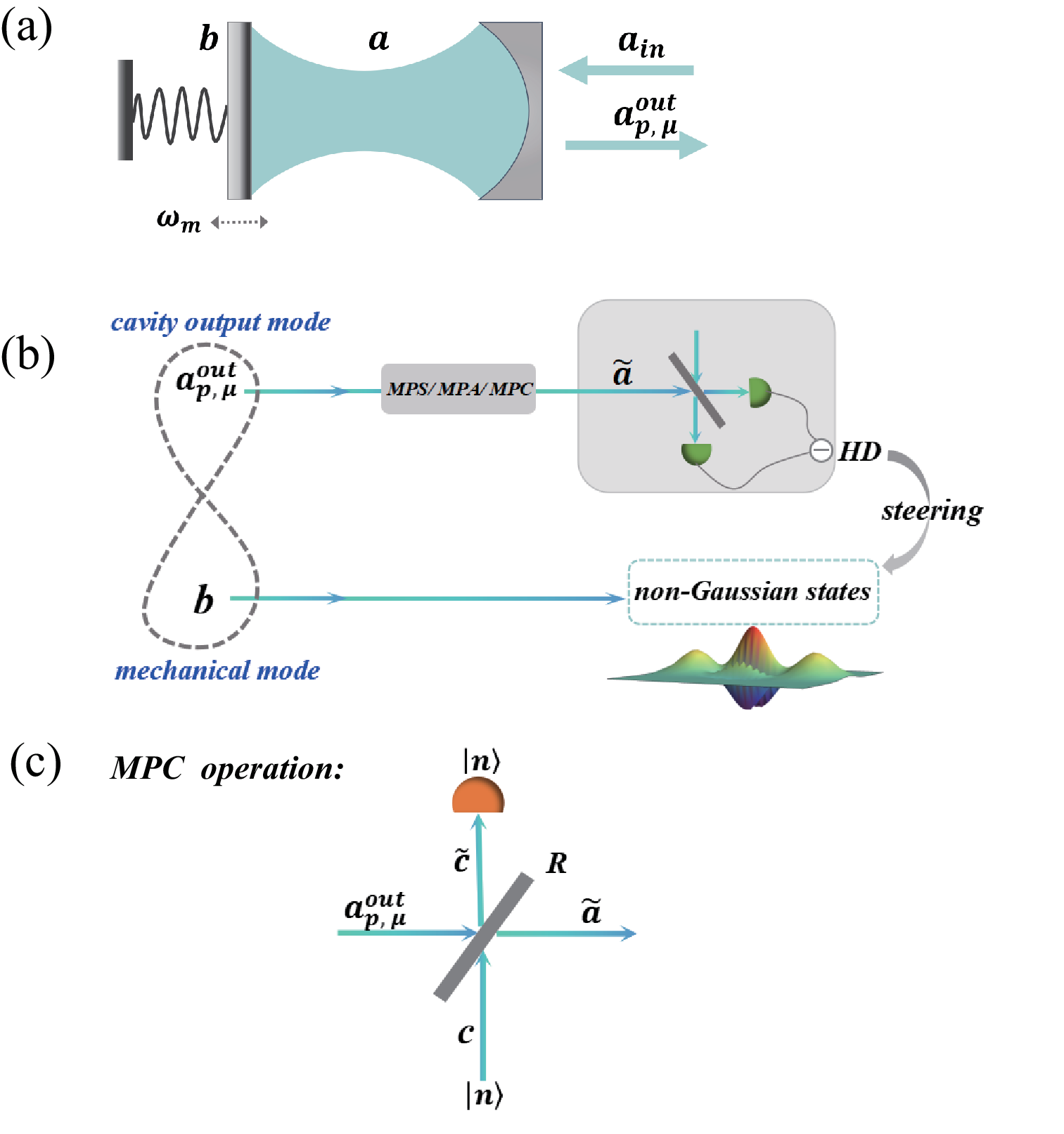}}}
\vspace{0cm}
\caption{(a) Schematic plot of a generic cavity optomechanical system in which the cavity field is dispersively coupled to a mechanical oscillator (denoted by $b$) of frequency $\omega_m$ and driven by a pulsed or continuous lasers. (b) Illustration of generating non-Guassian OM steerable states via the multiphoton operation (MPS, MPA or MPC) on the output field from the OM cavity and remotely preparing mechanical Sch\"{o}rdinger cat states via the subsequent homodyne detection (HD). (c) MPC operation: the output field $a_{p,\mu}^{out}$ from the OM cavity and an ancillary mode $c$ prepared in an $n$-photon Fock state, are combined at a beam splitter with reflectivity $R=\sin^{2}\theta$, and then the field from the outport $\tilde{c}$ of the beam splitter is detected on Fock state with the same $n$ photons.}
\label{fig2}
\end{figure}

\section{Cavity Optomechanical Systems}
We consider a generic cavity OM system in which the cavity and mechanical oscillator are dispersively coupled to each other via radiation pressure inside the cavity driven by a laser of frequency $\omega_p$, as shown in Fig.\ref{fig2} (a). The nonlinear OM interaction is described by the Hamiltonian ($\hbar=1$)
\begin{align}
\hat{H}=\delta\hat{A}^\dagger\hat{A}+\omega_m\hat{B}^\dagger
\hat{B}+g_0\hat{A}^\dagger\hat{A}(\hat{B}^\dagger+\hat{B})
+i(\varepsilon\hat{A}^\dagger-\varepsilon^*\hat{A}),
\label{eh}
\end{align}
where the annihilation operator $\hat{A}~(\hat{B})$ denotes the cavity (mechanical) mode of resonant frequency $\omega_c~(\omega_m)$, $g_0$ is the single-photon OM coupling strength,
the detuning $\delta = \omega_c - \omega_p$, and $\varepsilon$ is the driving amplitude. With the Hamiltonian (\ref{eh}), the average values $\alpha\equiv \langle\hat A\rangle$ and $\beta\equiv \langle\hat B\rangle$ can be determined by the equations
\begin{align}
\frac{d}{dt}\alpha&=-(\kappa_c+i\delta)\alpha-ig_0\alpha(\beta+\beta^*)+\varepsilon,\\
\frac{d}{dt}\beta&=-(\gamma_m+i\omega_m)\beta-ig_0|\alpha|^2,
\end{align}
where $\kappa_c~(\gamma_m)$ is the cavity (mechanical) loss rate. The steady-state values of $\alpha$ and $\beta$ can therefore be obtained as $\alpha_s = |\varepsilon|/\sqrt{\kappa_c^{2}+\Delta^{2}}$, $\beta_s = -ig_0\alpha_s^2 /(\gamma_m + i\omega_m)$, with  $\Delta = \delta + 2g_0Re(\beta_s)$.

By expressing the operators $\hat{A}$ and $\hat{B}$ as $\hat{A} = \alpha_s + \hat{a}$ and $\hat{B} = \beta_s + \hat{b}$, where $\hat{a}~(\hat{b})$ denote  quantum fluctuations of the cavity (mechanical) mode, for strong driving field such that $\alpha_s^2\gg \langle\hat{a}^\dag\hat{a}\rangle$ and $|\beta_s|^2\gg \langle\hat{b}^\dag\hat{b}\rangle$, the Hamiltonian (13) can be linearized around the steady-state amplitudes as
\begin{align}
\hat{H}_{lin} = \Delta\hat{a}^\dagger\hat{a}+\omega_m\hat{b}^\dagger\hat{b}
+g(\hat{a}^\dagger+\hat{a})(\hat{b}^\dagger+\hat{b}),
\label{e14}
\end{align}
where the collective OM coupling $g = g_0\alpha_s$. The quantum Langevin equations for the operators $\hat{a}$ and $\hat{b}$ are satisfied by
\begin{subequations}
\begin{align}
\frac{d}{dt}\hat{a}&=-(\kappa_c+i\Delta) \hat{a}-ig(\hat b+\hat{b}^\dagger)+\sqrt{2\kappa _c}\hat{a}^{in}(t),\\
\frac{d}{dt}\hat{b}&=-(\gamma_m+i\omega_m)\hat{b}-ig(\hat a+\hat{a}^\dagger)+\sqrt{2\gamma _m}\hat{b}^{in}(t),
\end{align}
\label{Le}
\end{subequations}
where $\hat{a}^{in}(t)$ and $\hat{b}^{in}(t)$, respectively, represent vacuum noise entering the cavity and thermal fluctuations from the environment of the mechanical oscillator, with the nonzero correlations $\langle\hat{a}^{in}(t)\hat{a}^{in\dagger}(t')\rangle=\delta(t-t')$, $\langle\hat{b}^{in\dagger}(t)\hat{b}^{in}(t')\rangle=\bar{n}_{th}\delta(t-t')$, and $\langle\hat{b}^{in}(t)\hat{b}^{in\dagger}(t')\rangle=(\bar{n}_{th}+1)\delta(t-t')$. Here $\bar{n}_{th}=(e^{\hbar\omega_m/\kappa_B T}-1)^{-1}$ is the mean thermal excitation number of the mechanical thermal environment at temperature $T$, with $\kappa_B$ being the Boltzmann constant. In the following, we consider two situations in which the cavity is respectively driven by pulsed and continuous lasers.

\subsection{Pulsed drive}
We at first consider the cavity is driven by a blue-detuned ($\Delta=-\omega_m$) laser pulse with duration $\tau_w$. We assume that the pulse has an approximate constant amplitude, as considered in Refs.\cite{GS,KS}. We further consider the situation that the pulse duration $\tau_w\gg\kappa_c^{-1}$ and the mechanical frequency  $\omega_m\gg\{g,\kappa_c,\gamma_m\}$, the Hamiltonian of Eq.(\ref{e14}) is  also be applicable\cite{GS} and can be approximated as
\begin{align}
\hat{H}_{lin}\approx g(\hat{a}\hat{b}+\hat{a}^\dagger\hat{b}^\dagger),
\label{pdc}
\end{align}
under the rotating-wave approximation. Eq.(\ref{pdc}) describes the two-mode squeezing interaction between photons and phonons. We consider that
the pulse duration $\tau_w\ll (\gamma_m\bar{n}_{th} )^{-1}$, which allows us to neglect the mechanical damping and thus assures coherent dynamics over the full
duration. For the cavity dissipation rate  $\kappa_c \gg g$, the cavity mode can be adiabatically eliminated and we have $\hat{a}(t)\simeq \frac{g}{\kappa_c}\hat{b}+\sqrt{\frac{2}{\kappa_c}}\hat{a}^{in}(t)$ and $\dot{\hat{b}}\simeq G\hat{b}+\sqrt{2G}\hat{a}^{in\dagger}(t)$, with $G=g^2/\kappa_c$. By integrating this equation up to the time $\tau_w$ and defining $\hat{ a}_p^{in}=\sqrt{\frac{2G}{1-e^{2G\tau_\omega}}}\int_0^{\tau_\omega}\hat{a}^{in}(s)e^{-Gs}ds$, we have
\begin{align}
\hat{b}_p^{out}=\cosh{r}\hat{b}_p^{in}+\sinh{r}\hat{a}_p^{in\dagger},
\label{t1}
\end{align}
where the mechanical output mode $\hat{ b}_p^{out}=\hat{b}(\tau_\omega)$ and $\hat{ b}_p^{in}=\hat{b}(0)$, and $r=\cosh^{-1}{(e^{G\tau_w})}$. Similarly, by using the cavity input-output relation $\hat{a}^{out}=\sqrt{2\kappa_c}\hat{a}-\hat{a}^{in}$ and defining the cavity output operator $\hat{a}_p^{out}=\sqrt{\frac{2G}{e^{2G\tau_w}-1}}\int_0^{\tau_w}\hat{a}^{out}(s)e^{Gs}ds$,
we have
\begin{align}
\hat{a}_p^{out}=\cosh{r}\hat{a}_p^{in}+\sinh{r}\hat{b}_p^{in\dagger}.
\label{t2}
\end{align}
Eqs. (\ref{t1}) and (\ref{t2}) correspond to a two-mode squeezing transformation characterized by the operator $\hat{S}_p(r)=e^{r(\hat{a}_p^{in\dagger}\hat{b}_p^{in\dagger}
-\hat{a}_p^{in}\hat{b}_p^{in})}$. This means that after the interaction time $\tau_w$, the mechanical oscillator and the cavity output field are prepared in a two-mode squeezed state, i.e.,
\begin{align}
\hat{\rho}_{\hat{a}_p^{out}\hat{b}_p^{out}}
=\hat{S}_p(r)\hat{\rho}_{\hat{a}_p^{in}}
\otimes\hat{\rho}_{\hat{b}_p^{in}}\hat{S}_p^\dagger(r),
\label{sq}
\end{align}
where $\hat{\rho}_{\hat{a}_p^{in}}$ and $\hat{\rho}_{\hat{b}_p^{in}}$ denote the initial states of the optical cavity field and mechanical oscillator before the pulse entering the cavity.  The initial states are considered to be vacuum and thermal states with the mean thermal number $\bar n_0$ for the cavity field and mechanical oscillator, respectively. Ideally, when $\bar n_0\simeq0$ via ground-state cooling \cite{COMPP3}, the cavity field and mechanical oscillator are prepared in a standard two-mode squeezed vacuum. Thus, the output field is entangled with the mechanical oscillator with the blue-detuned laser pulse, as experimentally realized in Ref.\cite{KS}. Therefore, to ensure strong entanglement (i.e. large squeezing parameter $r$) between the mechanical oscillator and the output field, it is necessary for $g\tau_w\gg1$, since $G\tau_w\equiv \frac{g}{\kappa_c}(g\tau_w)$ and $\frac{g}{\kappa_c}$ needs to
be small. This can be achieved by adjusting the pulse duration. For instance, for the cavity loss rate $\kappa_c\simeq0.1\omega_m$ and the OM coupling $g\simeq0.1\kappa_c$, to achieve the squeezing parameter $r=1$, the duration $\tau_w\simeq430\omega_m^{-1}\ll (\gamma_m\bar{n}_{th} )^{-1}$ for a mechanical oscillator cooled to the near ground state  ($\bar{n}_{th}<1$) with the mechanical quality factor $Q\equiv \frac{\omega_m}{\gamma_m}\gtrsim 5\times 10^3$.

\subsection{Continuous drive}
We next consider that case that the cavity is continuously driven. To study the quantum correlations between the mechanical oscillator and the output field which consists of a continuum of modes, we define a temporal mode $\mu(t)$ from the output field $a^{out}(t)$, denoted by the annihilation operator
\begin{align}
\hat a_{\mu}^{out}=\int \mu^*(t')\hat a^{out}(t')dt',
\label{eq2}
\end{align}
which satisfies the commutation relation $[\hat a_{\mu}^{out}, \hat a_{\mu}^{out\dag}]=1$, leading to $\int |\mu(t)|^2dt=1$.
The filtering of the temporal mode $\hat a_{\mu}^{out}$ from the output field $\hat a^{out}(t)$ can be viewed as the undirectionally injecting the output field into a a virtual cavity (filter), which can be characterized by coupling the OM cavity field cascadedly to the virtual cavity. This virtual cavity field thus acts as the filtered field from the output field $\hat a^{out}(t)$, with the desired temporal mode function $\mu (t)$ which determines the cascaded coupling strength $g_{\mu}(t)$ from the OM cavity to the virtual cavity, given by \cite{Alex,Alexander} (also see the Appendix)
\begin{align}
g_{\mu}(t)=-\frac{\mu^*(t)}{\sqrt{\int_0^{t} dt'|\mu(t')|^2}}.
\label{eq3}
\end{align}
We consider the constant coupling, i.e., $g_{\mu}=\sqrt{{\kappa_\mu}}$, for simplicity.
In this description, the master equation for the whole cascaded system including the cavity field $\hat a$, mechanical mode $\hat b$, and the corresponding virtual cavity field $\hat a_{\mu}^{out}$ can be described by the master equation
\begin{align}
\frac{d\hat \rho}{dt}=&-i[\hat{H}_{lin}+\hat H_{ex},\hat \rho]+\mathcal L[\hat J]\hat \rho\nonumber\\
&+\mathcal L[\sqrt{\gamma_m(\bar{n}_{th}+1)}\hat{b}]\hat\rho+\mathcal L[\sqrt{\gamma_m\bar{n}_{th}}\hat{b}^\dag]\hat \rho,
\label{me}
\end{align}
where the unidirectional-coupling resulted coherent exchange coupling
\begin{equation}
\begin{split}
\hat H_{ex}&=\frac{i}{2}(\sqrt{\kappa_c\kappa_\mu}\hat a^\dag\hat a_{\mu}^{out}-h.c.),
\end{split}
\end{equation}
and the collective decay $\mathcal L[\hat o]\hat \rho=\hat o\hat \rho\hat o^\dag-\frac{1}{2}(\hat o^\dag\hat o\hat \rho+\hat \rho\hat o^\dag\hat o)$, with the collective operator
\begin{align}
\hat J=\sqrt{\kappa_c}\hat a+\sqrt{\kappa_\mu}\hat a_{\mu}^{out},
\end{align}
describing the collective dissipation of the OM cavity and virtual cavity fields into the common vacuum reservoir. In the following section, the master equation (\ref{me}) will be numerically solved, and we are interested in the regime of steady states for the case of continuous drive. With the density operator $\hat \rho$, the reduced density operator $\hat\rho_{\hat a_{\mu}^{out}\hat b}$ describing the two-mode state of the mechanical oscillator and the filtered
cavity output field can be obtained via $\hat\rho_{\hat a_{\mu}^{out}\hat b}=\text{Tr}_a[\hat \rho]$.

\section{non-Gaussian OM steering via multiphoton operations}
\subsection{OM non-Gaussian states}
After obtaining the density matrices $\rho_{\hat a^{out}\hat b}= \{\hat{\rho}_{\hat{a}_p^{out}\hat{b}_p^{out}}, \hat\rho_{\hat a_{\mu}^{out}\hat b}\}$ of the two-mode OM system respectively driven by pulsed and continuous lasers, we proceed to consider the generation of non-Gaussian OM states via three non-Gaussian operations: MPS, MPA, and MPC on the output field ($\hat a_p^{out}$ or $\hat a_\mu^{out}$) leaking from the OM cavity, as shown in Fig.2(b). The output states described by the density operators of the mechanics and output field after the multiphoton operations can be expressed as \begin{subequations}
\begin{align}
\hat\rho_{\tilde{a}b}^{MPS}&=\hat{a}_o^n\hat\rho_{\hat a^{out}\hat b}\hat{a}_o^{\dag n},\\
\hat\rho_{\tilde{a}b}^{MPA}&=\hat{a}_o^{\dag n}\hat\rho_{\hat a^{out}\hat b}\hat{a}_o^n,\\
\hat\rho_{\tilde{a}b}^{MPC}&=\hat{B}^n\hat\rho_{\hat a^{out}\hat b}\hat{B}^{\dag n},
\end{align}
\end{subequations}
where $\hat a_o=\{\hat a_p^{out}, \hat a_\mu^{out}\}$. Specifically, the MPC operation as shown in Fig.2(c) describes the process in which the output field $\hat a_p^{out}$ or $\hat a_\mu^{out}$ from the OM cavity and an ancillary mode, described by the bosonic operator $\hat c$ and prepared in an $n$-photon Fock state, are combined at a beam splitter with reflectivity $R=\sin^{2}\theta$, and then one of the fields from the outports of the beam splitter is detected on Fock state with the same $n$ photons \cite{cata}. A catalytic operator of $n$ photons is defined as
\begin{equation}
\begin{split}
\hat B^{n}&={\langle n}|\hat B(\theta)|n\rangle={\langle n}|\hat B(\theta)\sum\limits_{\emph{l}_a=0}|n\rangle|\emph{l}_a\rangle\langle\emph{l}_a|
\nonumber\\
&=\sum\limits_{\emph{l}_a=0} B_{n,l_a}|\emph{l}_a\rangle\langle\emph{l}_a|,
\end{split}
\end{equation}
where the beam splitter operator $\hat B(\theta)=\exp\{\theta(\hat c \hat a_o^{\dagger}- \hat c^{\dagger} \hat a_o)\}$ and
\begin{align}
B_{n,l_a}=&\sum\limits^{n}_{i=0}(-1)^{n-i}
\mathrm{C}_n^i\mathrm{C}_{\emph{l}_a}^{n-i}(\cos\theta)^{\emph{l}_a+2i-n}
(\sin\theta)^{2n-2i}.
\end{align}

Following the non-Gaussian operation on the mode $a$, we further consider the homodyne detection on the quadrature $\hat q_{\tilde{a}}=(\hat {\tilde{a}}+\hat {\tilde{a}}^\dag)/\sqrt{2}$ of the output mode ${\tilde{a}}$ from the beam splitter. Conditioned on the detection outcomes $q_{\tilde{a}}$, the density operator of the mode $b$
\begin{align}
\hat \rho_b(q_{\tilde{a}})=\hat {\tilde{\rho}}_b(q_{\tilde{a}})/\text{Tr}_b \big[\hat {\tilde{\rho}}_b(q_{\tilde{a}})\big],
\label{e3}
\end{align}
where the unnormalzied operator $\hat {\tilde{\rho}}_b(q_{\tilde{a}})=\text{Tr}_{{\tilde{a}}} \big[(\hat {\mathcal M}_{{\tilde{a}}} \otimes \hat I_b)\hat\rho_{\tilde a b}(\hat I_b\otimes\hat {\mathcal M}_{{\tilde{a}}})\big]$ and the projection operator $\hat {\mathcal M}_{{\tilde{a}}}=|q_{\tilde{a}}\rangle\langle q_{\tilde{a}}|$, which can be calculated in the Fock space $\{|n_{\tilde a}\rangle\}$ with $\langle q_{\tilde a}\mid n_{\tilde a}\rangle=\frac{1}{\pi^{1/4}}\frac{1}{\sqrt{2^{n_{\tilde a}}n_{\tilde a}!}}e^{-q_{{\tilde a}}^{2}/2}H_{n_{\tilde a}}(q_{\tilde a})$, $H_{n_{\tilde a}}$ the Hermite polynomial of order $n_{\tilde a}$.

Specially, for the OM states prepared in a two-mode squeezed vacuum state of Eq.(\ref{sq}) with the pulsed drive, the conditioned states $\hat \rho_b(q_{\tilde{a}})$ of the mechanical oscillator can be analytically derived as
\begin{align}
\hat\rho^{MPS}_b(q_{\tilde{a}})=& N_{MPS}\sum\limits_{\emph{l}_b(\emph{l}'_b)=n}\frac{H_{\emph{l}_b-n}(q_{{\tilde{a}}})
H_{\emph{l}'_b-n}(q_{{\tilde{a}}})e^{-q_{{\tilde{a}}}^{2}}}{\pi^{\frac{1}{2}}2^{\frac{\emph{l}_b+\emph{l}'_b-2n}{2}\sqrt{(\emph{l}_b-n)!(\emph{l}'_b-n)!}}}\nonumber\\
&\sqrt{\emph{l}_b\emph{l}'_b}^{n}(\sech r)^{2}(\tanh r)^{\emph{l}_b+\emph{l}'_b}|\emph{l}_b\rangle\langle\emph{l}'_b|,
\label{rms1}
\end{align}
\begin{align}
\hat\rho^{MPA}_b(q_{\tilde{a}})=& N_{MPA}\sum\limits_{\emph{l}_b(\emph{l}'_b)=0}\frac{H_{\emph{l}_b+n}(q_{\tilde{a}})
H_{\emph{l}'_b+n}(q_{\tilde{a}})e^{-q_{\tilde{a}}^{2}}}{\pi^{\frac{1}{2}}2^{\frac{\emph{l}_b+\emph{l}'_b+2n}{2}\sqrt{(\emph{l}_b+n)!(\emph{l}'_b+n)!}}}\nonumber\\
&\sqrt{(\emph{l}_b+1)(\emph{l}'_b+1)}^{n}(\sech r)^{2}(\tanh r)^{\emph{l}_b+\emph{l}'_b}|\emph{l}_b\rangle\langle\emph{l}'_b|,
\label{rms2}
\end{align}
\begin{align}
\hat\rho^{MPC}_b(q_{\tilde{a}})=& N_{MPC}\sum\limits_{\emph{l}_b(\emph{l}'_b)=0}\frac{H_{\emph{l}_b+n}(q_{{\tilde{a}}})
H_{\emph{l}'_b+n}(q_{{\tilde{a}}})e^{-q_{{\tilde{a}}}^{2}}}{\pi^{\frac{1}{2}}2^{\frac{\emph{l}_b+\emph{l}'_b+2n}{2}\sqrt{(\emph{l}_b+n)!(\emph{l}'_b\pm n)!}}}\nonumber\\
&B_{n,\emph{l}_b}B_{n,\emph{l}'_b}(\sech r)^{2}(\tanh r)^{\emph{l}_b+\emph{l}'_b}|\emph{l}_b\rangle\langle\emph{l}'_b|,
\label{rms3}
\end{align}
where $N_{MPS}$, $N_{MPA}$, and $N_{MPC}$ are the normalization factors. With the conditional states (\ref{e3}) of the mechanics, the metrology-based steering criteria in Eq.(\ref{e13}) can be calculated.
\begin{figure*}[t]
\vspace{0cm}
\centerline{\hspace{0cm}\scalebox{0.38}{\includegraphics{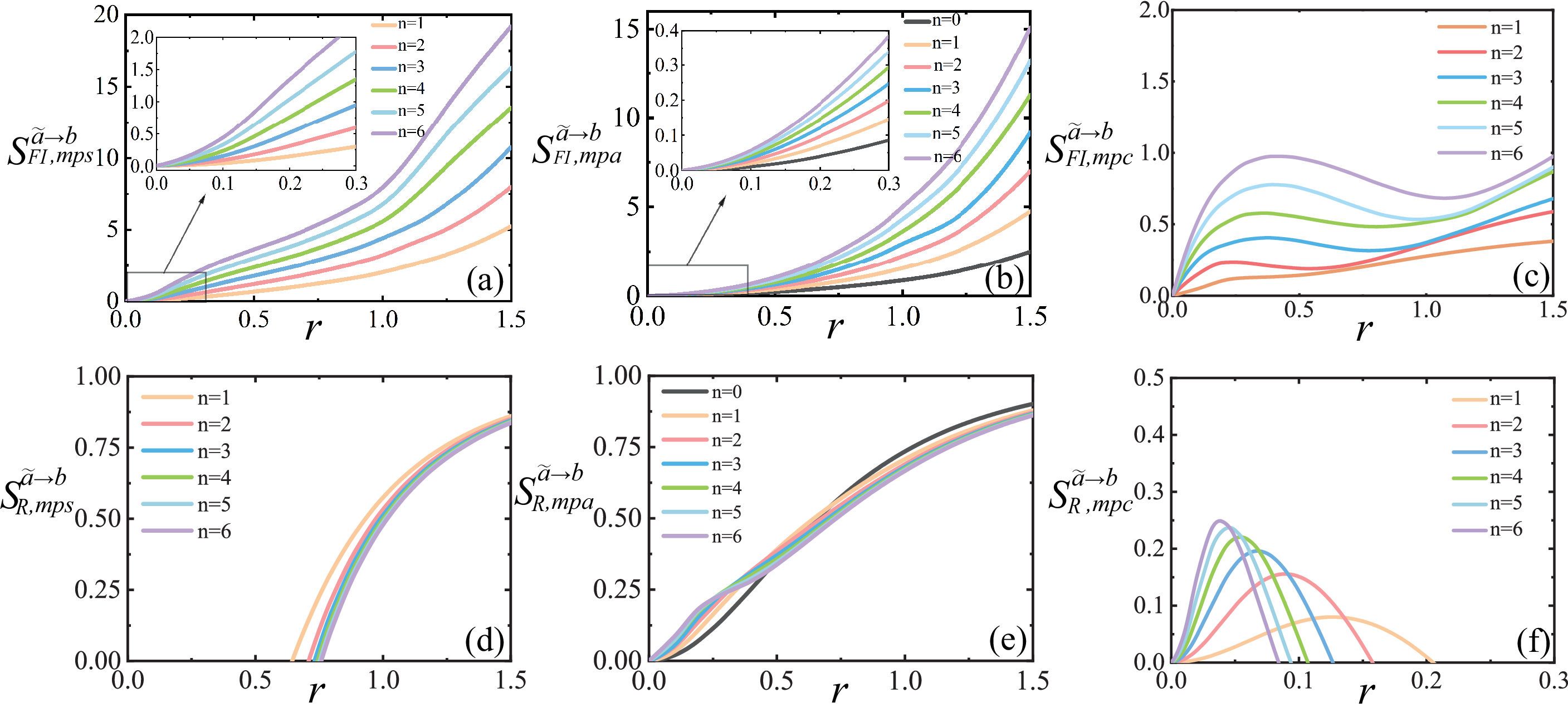}}}
\vspace{0cm}
\caption{The dependences of the FI-based steering [(a)-(c)] and Reid's variance-based steering [(d)-(f)] between the mechanical oscillator and cavity output field on the squeezing parameter $r$ in the pulse-drive case with the MPS, MPA and MPC operations and initial vacuum states, for different photon number $n$ in the multiphoton operations. The beam-splitter transmissivity $T=1-R=0.1$ in the MPC (similarly hereinafter, unless otherwise specified).}
\label{fig3}
\end{figure*}

\subsection{Results for non-Gaussian OM steering}
We first study the OM steering for the case of pulsed drive. By fixing  $\theta_A=0$ and $\theta_B=\pi/2$, the steering from the output field to the mechanical oscillator after the non-Gaussian multiphoton operations is plotted in Fig.\ref{fig3}. It shows from Fig.\ref{fig3} (a) and (b) that the steering, witnessed by the FI-based criterion,  increases with the photon number $n$ involved in the MPS and MPA operations, for the given squeezing parameter $r$. This means that for a two-mode squeezed vacuum state,  subtracting or adding more photons to one mode leads to a larger value of the quantum steering exhibited in the state. At the same time, the steering increases monotonically with the squeezing parameter $r$, for the fixed photon number $n$. This monotonicity arises from the fact that the entanglement of a two-mode squeezed vacuum state increases monotonically with the squeezing parameter. In addition, for the same squeezing $r$ and photon number $n$, the steering enhancement induced by the MPS is larger than that resulted from the MPA. The reason may be that the MPS on the optical mode effectively reduces the photon number distribution of the OM two-mode squeezed vacuum in Fock-state space, and thus the MPS operation results in the mechanical mode exhibiting smaller fluctuations (excluding the vacuum state) and enables easier steering of the mechanical states, compared to the MPA operation.

\begin{figure}[t]
\vspace{0cm}
\centerline{\hspace{0cm}\scalebox{0.33}{\includegraphics{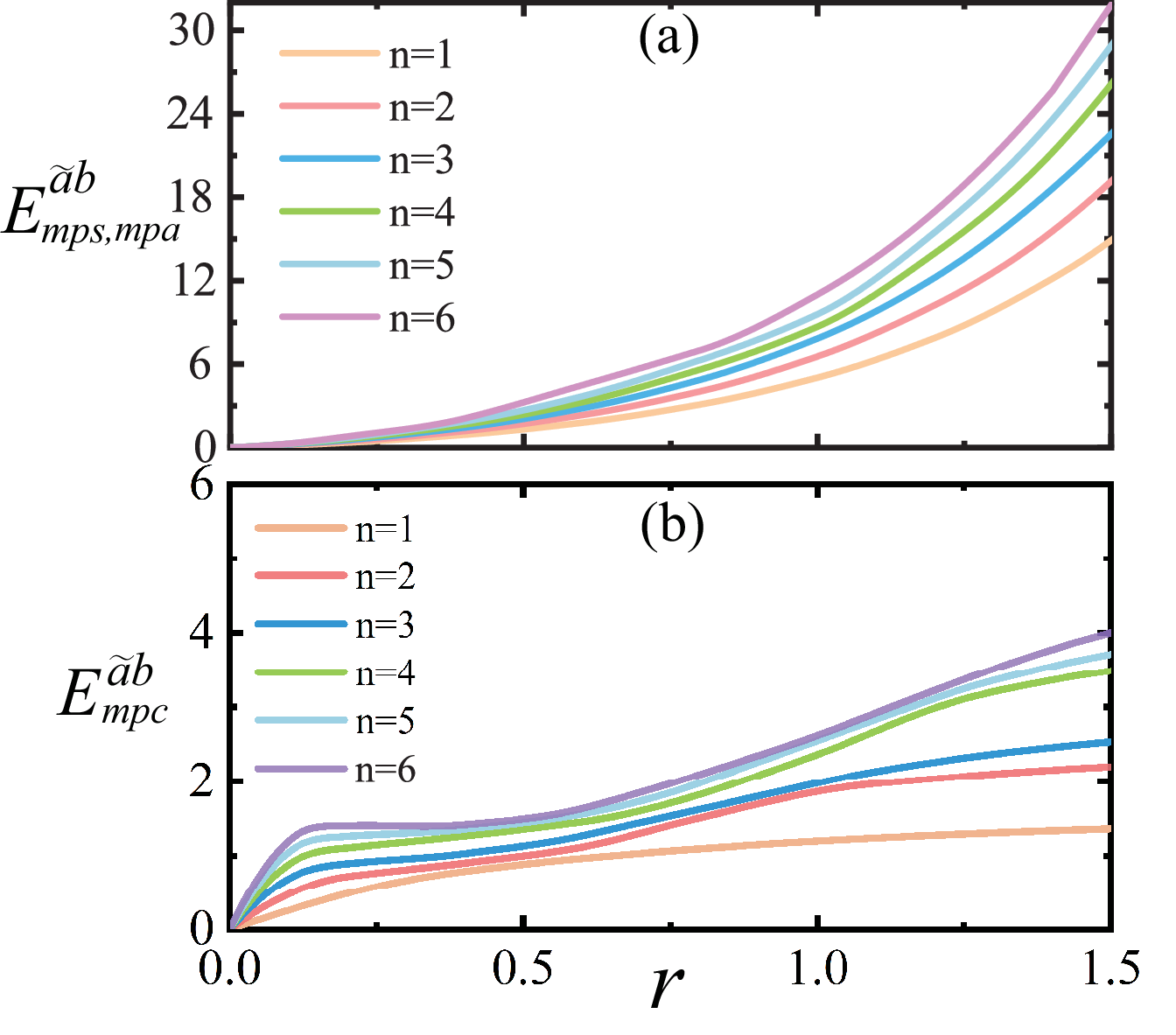}}}
\vspace{0cm}
\caption{The dependences of the OM entanglement on the squeezing parameter $r$ with the MPS and MPA [(a)] and MPC [(b)] operations for initial vacuum states and different photon number $n$.}
\label{fig4}
\end{figure}

\begin{figure*}[t]
\vspace{0cm}
\centerline{\hspace{0cm}\scalebox{0.42}{\includegraphics{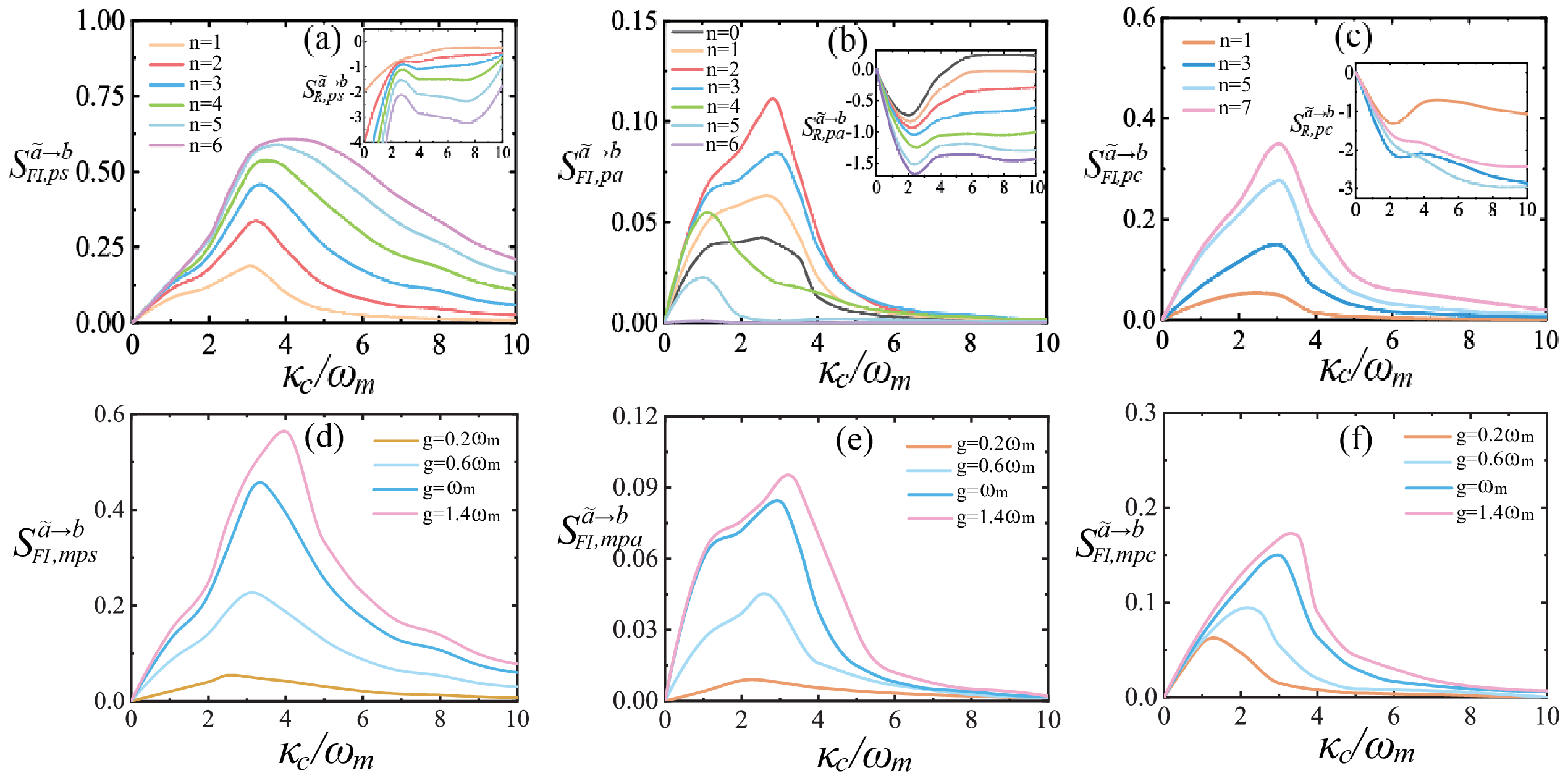}}}
\vspace{0cm}
\caption{The dependences of the FI-based steering and Reid's variance-based steering (inserts) between the mechanical oscillator and the cavity output field on $\kappa_c/\omega_m$ in the case of continuous drive with MPS, MPA, and MPC operations. The other parameters $\Delta=\omega_m$, $g=\omega_m$ [(a)-(c)], $n=3~[(d)-(f)]$, $\kappa_{\mu}=64\omega_m$, $\gamma_m=0$, and $\bar{n}_{th}=0$.}
\label{fig5}
\end{figure*}

For comparison, we plot correspondingly in Fig.\ref{fig3} (d)-(f) the steering characterized by the Reid's variance-based steering criterion defined by \cite{SR}
\begin{align}
S_R^{\tilde{a}\rightarrow b}=1-\frac{2\sqrt{Var^{b\mid\tilde{a}}(\hat q_b)Var^{b\mid\tilde{a}}(\hat p_b)}}{\big[\hat q_{b},\hat p_{b}\big]}>0.
\end{align}
Here $Var^{b\mid\tilde{a}}(\hat {\mathcal O})=\int o_b^2 P_{o_{\tilde{a}}}^b (o_b) d o_b
-(\int o_b P_{o_{\tilde{a}}}^b(o_b)do_b)^2$ and $o=q, p$.
It is the inferred variance of the operator $\hat {\mathcal O}=\big\{\hat q_{b},\hat p_{b}\big\}$ for the two-mode states $\hat \rho_{b}^{\rm con}(q_{\tilde{a}})$ and $\hat \rho_{b}^{\rm con}(p_{\tilde{a}})$ conditioned on the outcomes $q_{\tilde{a}}$ and $p_{\tilde{a}}$ of the corresponding homodyne detection on $\hat q_{\tilde{a}}$ and $\hat p_{\tilde{a}}$, respectively.
We see that for the weak and moderate degrees of squeezing, the Reid's criterion fails to detect the non-Gaussian steering induced by the MPS. While for the larger values of the squeezing ($r\gtrsim 0.53$), the steering with the MPS operation is present but it slightly decreases with the photon number $n$ for the given squeezing, in contrast to the steering revealed by the FI-based criterion. For the case of the MPA operation, the steering witnessed by the Reid's criterion is present for the whole range of the squeezing. For small values of squeezing, the steering increases with the photon number
$n$, but as the squeezing continues to increase, the steering decreases with
$n$. This is also different from the steering properties revealed with the FI-based criterion. To better compare the two steering criteria,  we consider the bipartite entanglement between the mechanical oscillator and the cavity output field, which can be sufficiently characterizes by the negativity $E(\hat\rho_{\tilde{a}b})$ defined as
\begin{align}
E^{\tilde{a}b}=\frac{\|\hat{\rho}_{\tilde{a}b}^{T_b}\|-1}{2},
\end{align}
where $\hat{\rho}_{\tilde{a}b}^{T_b}$ is the partial transpose over the mechanical mode and
$\|\hat o\|=\text{Tr}(\sqrt{\hat o^\dagger\hat o})$ is the trace norm.
As shown in Fig.\ref{fig4}, the OM entanglement increases with the increase of the squeezing parameter and the photon number $n$ in both cases of MPS and MPA operations, similarly to the features of the steering characterized by the FI-based criterion. Note that different from the steering, the non-Gaussian entanglement revealed by the negativity via the partial transpose is the same for both MPS and MPA operations.

In Fig.\ref{fig3} (c) and (f), the non-Gaussian steering characterized respectively by the FI-based and Reid's criteria is respectively plotted for the case of MPC operation. It is clearly shown that the steering features revealed by the two criteria are also quite different. The Reid's criterion shows the steering just appears in the presence of very small squeezing $r$, but the FI-based criterion reveals that the steering exists with nonzero squeezing and it also increases with the photon number $n$ for the given squeezing. The OM entanglement in this case is plotted in Fig.\ref{fig4} (b) and we see that it has the similar dependences, to the FI-based steering, on the squeezing parameter $r$ and photon number $n$ under the MPC operation. In addition, we see from Fig.\ref{fig3} that for large squeezing $r$, the non-Gaussian steering via the MPS and MPA operations is much stronger than that by the MPC operation, with the same photon number $n$. Therefore, indicated from the dependence of the entanglement on the photon number $n$ in the MPS, MPA, and MPC operations, it can be concluded that the FI-based criterion is more efficient for witnessing non-Gaussian quantum steering than the Reid's criterion. The reason is that the Reid's criterion involves the conditional variances of linear operators, and since the change in variances is not significant when the photon number $n$ in the multiphoton operations changes only slightly, this leads to subtle changes in the steering based on the Reid's criterion.

We next consider the non-Gaussian OM steering for the case of continuous drive.
In Fig.\ref{fig5}, we plot the effects of the cavity loss rate $\kappa_c$ and the OM coupling $g$ on the light-to-mechanics steering when the cavity output field is subject to MPS, MPA, and MPC operations.
It clearly shows from Fig.\ref{fig5}(a)-(c) that the non-Gaussian steering increases at first and then decreases with the increasing of the cavity dissipation rate $\kappa_c$ and moreover the optimal steering is achieved for $\kappa_c>\omega_m$, i.e., in the sideband-unresolved regime, in contrast to the case of pulsed drive which operates in the sideband-resolved regime. This behavior can be explained with the cascade master equation (\ref{me}), which shows that the increase of the cavity dissipation enhances the coupling between the mechanical oscillator and the filtered output field (the virtual cavity's mode), leading to stronger steering. As the dissipation continues to increase, the decoherence dominates, resulting in a decrease of the steering. As shown in Fig.5 (d)-(f), the steering increases with the OM coupling strength. This is also clearly due to the fact that the increase in the coupling enhances the OM entanglement, which in turn lead to stronger steering via the non-Gaussian operations.

Within the parameter range for the pulse-drive case (i.e.,  $\omega_m\gg\{\kappa_c,g,\gamma_m\}$ and $g\ll\kappa_c$), we observe from Fig.3(a)-(c) and Fig.5 that for the same parameters, the achievable amount of non-Gaussian steering with the continuous drive is much weaker than that with the pulse drive. This is because, for the pulse-drive case, the mechanical oscillator and the output pulse field are prepared in a pure two-mode squeezed vacuum when the mechanical damping is negligible, i.e., for $\tau_w\ll (\gamma_m\bar {n}_{th})^{-1}$. Moreover, with the given parameters, we can still chose the pulse duration to achieve a large value of the squeezing parameter $r$, leading to strong OM steering. Although the two-mode squeezing OM interaction is more beneficial for the generation of the OM steering, as seen in the pulse case, the OM coupling strength $g$ with the continuous drive in the blue-detuned regime is severely limited by the stability condition, which merely brings about very weak OM steering in the steady-state regime. With stronger coupling and still maintain the stability, the system should be continuously driven in the red-detuned regime. This leads to the non-resonant two-mode squeezing interaction, e.g., $\hat{H}_{lin} =
g\hat{a}^\dagger\hat{b}+g\hat{a}^\dagger\hat{b}^\dagger e^{2i\omega_mt}+h.c.$ with the detuning $\Delta=\omega_m$. Therefore, the maximally achievable steering is decreasing with the other same parameters in the continuous-drive case.  The maximal steering achieved by the MPS and MPC operations increases with the photon number $n$ for a given dissipation $\kappa_c$, and the optimal value of $\kappa_c$ for achieving the maximum steering shifts to larger values as the photon number $n$ increases. In contrast, the maximal steering achieved via the MPA operation increases at first and then deceases as the photon number $n$ increases. By comparing Fig.\ref{fig3} and Fig.\ref{fig5}, it is evident that the MPS operation is much more efficient for achieving the non-Gaussian OM steering than the MPA operations. In addition, we see that the MPC can lead to stronger steering than the MPA, different from the case of pulsed drive. It is shown from the inserts in Fig.\ref{fig5} the Reid's criterion  fails to detect the non-Gaussian steering when operating MPS, MPA and MPC for the case of continuous drive. This shows again that the FI-based steering criterion is more effective for detecting the non-Gaussian steering.

In Fig.6, the effect of thermal mechanical fluctuations on the steerable OM correlations is evaluated. We consider the photon number $n=3$ in the multiphoton operations. As shown in Fig.6 (a) which plots the effect of initial thermal phonon number $\bar n_0$ of the mechanical oscillator for the case of pulsed drive, the non-Gaussian OM steering decreases as $\bar n_0$ increases and it can exist up to $\bar n_0\approx 2$ and then disappears as the initial mechanical thermal excitation increases. This means that the precooling of the mechanical oscillator to the ground states is necessary for the pulse-drive case. For mechanical resonators with frequencies in the GHz range, the excitation number can be neglected at millikelvin temperatures achieved using a dilution refrigerator \cite{gc}. Fig.6(b) plots the effect of the thermal excitation number $\bar n_{th}$ of the phononic environment in the case of continuous drive. We can see that the non-Gaussian OM steering with the selected cavity output component is robust against the thermal noise and can still exist around $\bar n_{th}\approx 10^4$ for the mechanical damping rate $\gamma_m=10^{-3}\omega_m$. As shown in the insert in Fig.6(b), the tolerance to thermal noise decreases as the damping rate increases.

\begin{figure}[H]
\vspace{0cm}
\centerline{\hspace{0cm}\scalebox{0.45}{\includegraphics{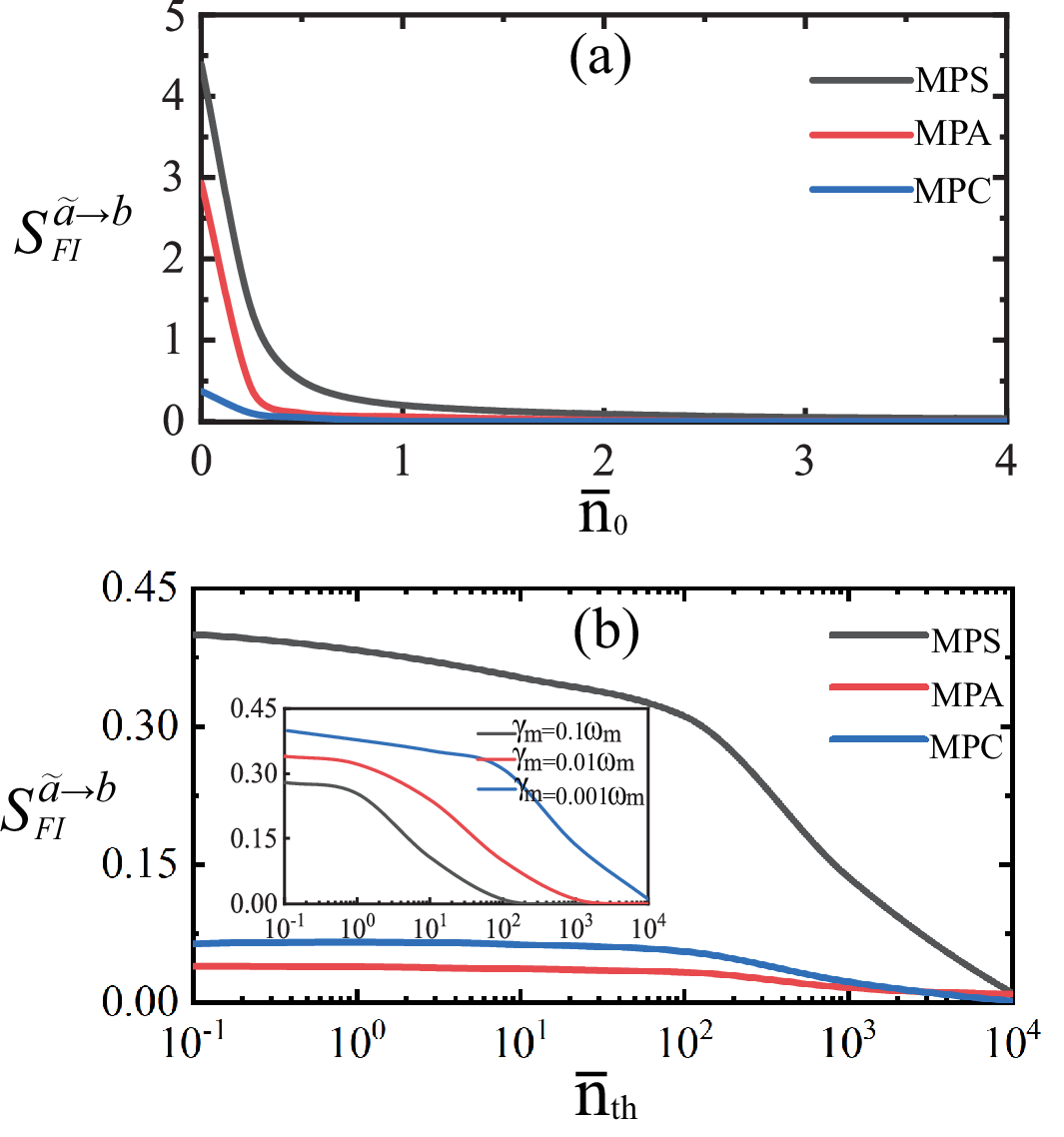}}}
\vspace{0cm}
\caption{The dependences of the OM steering for the photon number $n=3$ in the multiphoton operations on the initial thermal phonon number $\bar n_0$ in the pulse-drive case with the parameter $r=1$ (a) and on the mean thermal excitation number $\bar n_{th}$ in the continuous-drive case (b) with $\gamma_m=10^{-3}\omega_m$ and the other the parameters same as in Fig.(\ref{fig5}). The insert plots the effect of different mechanical damping rates.}
\label{fig6}
\end{figure}

\section{Remote preparation of mechanical large-size Sch\"{o}rdinger cat sates via homodyne detection}
From the above discussion, through the local non-Gaussian multiphoton operations on the cavity output field, the non-Gaussian OM steering can be achieved, which is verified via the homodyne detection with the FI-based steering witness within the metrological protocal. This homodyne detection projects the mechanical oscillator in non-Gaussian states $\hat\rho_b(q_{\tilde{a}})$ in Eqs.(\ref{rms1})-(\ref{rms3}) with the help of the non-Gaussian OM steerable correlations.
We now study the quantum features of these mechanical states. The genuine nonclasscality of the non-Gaussian states is embodied by the corresponding Wigner-function negativity \cite{N}, defined via
\begin{align}
\mathcal N\equiv\int\Big[\big|W_b(\beta)\big|-W_b(\beta)\Big]d^{2}\beta,
\end{align}
where the Wigner function $W_b(\beta)=\frac{1}{\pi^{2}}\int d^2\xi \text{Tr}\big[e^{\xi\hat b^\dag -\xi^*\hat b}\hat \rho_b(q_{\tilde a})\big] e^{(\beta\xi^{*}-\beta^{*}\xi)}$, with phase-space variable $\beta=q_b+ip_b$.
In addition, the fidelity of the state $\hat \rho_b(q_{\tilde a})$
 with respect to an ideal cat state $\hat \rho_\alpha=|\psi_\alpha\rangle_{\rm cat}\langle \psi_\alpha|$ with the amplitude $\alpha$ is
\begin{align}
F_\pm=\text{Tr}\Big[\sqrt{\sqrt{\hat \rho_\alpha}\hat \rho_b(q_{\tilde a})\sqrt{\hat \rho_\alpha}}\Big],
\end{align}
where $|\psi_\alpha\rangle_{\rm cat}=\frac{1}{\sqrt{2(1\pm e^{-2|\alpha|^{2}})}}\big(|\alpha \rangle\pm|-\alpha\rangle\big)$, with the sign $``+"~(-)$ denoting even (odd) cat states.

\begin{figure}[H]
\centerline{\hspace{0cm}\scalebox{0.24}{\includegraphics{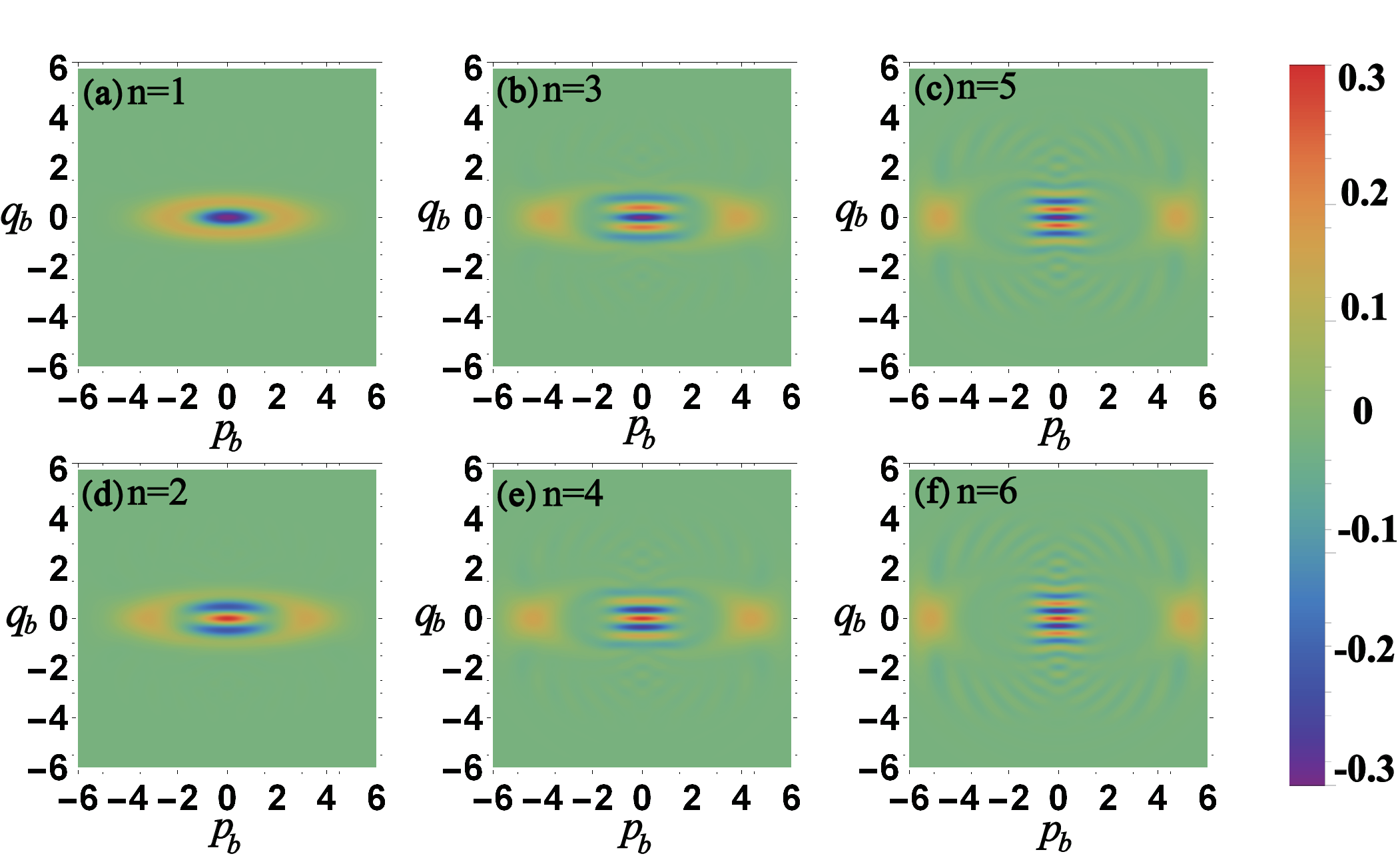}}}
\caption{The density plots of the Wigner function $W_b(q_b,p_b)$ of the conditioned state $\hat\rho_b(q_{\tilde a}=0)$ of the the mechanical oscillator in initial vacuum after subtracting $n$ photons from cavity output mode in the pulse-drive case, with the squeezing parameter $r=1$.}
\label{fig7}
\end{figure}

\begin{figure}[H]
\centerline{\hspace{0.3cm}\scalebox{0.24}{\includegraphics{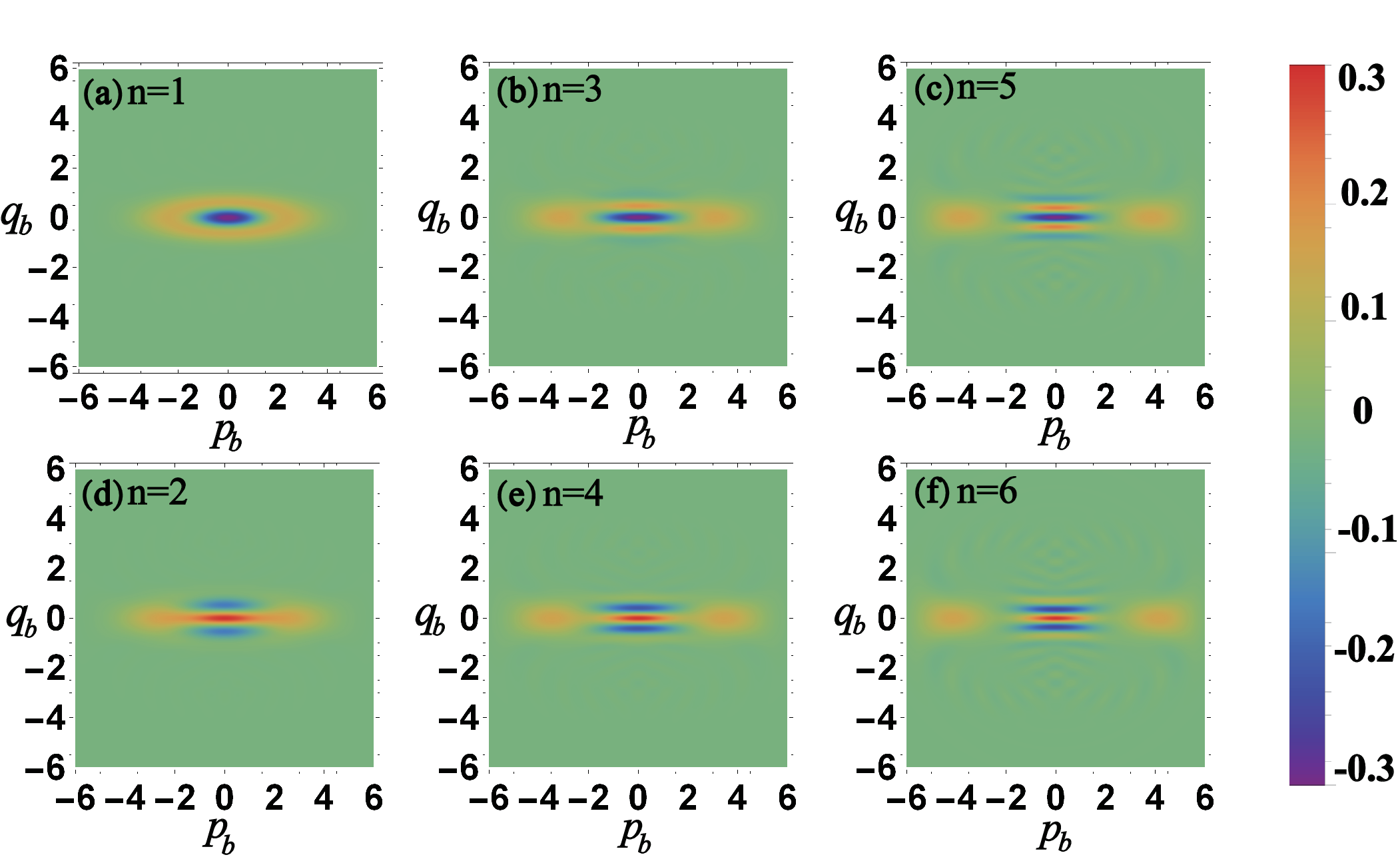}}}
\caption{The same as Fig.\ref{fig7} but for adding $n$ photons.}
\label{fig8}
\end{figure}

\begin{figure}[H]
\centerline{\hspace{0cm}\scalebox{0.27}{\includegraphics{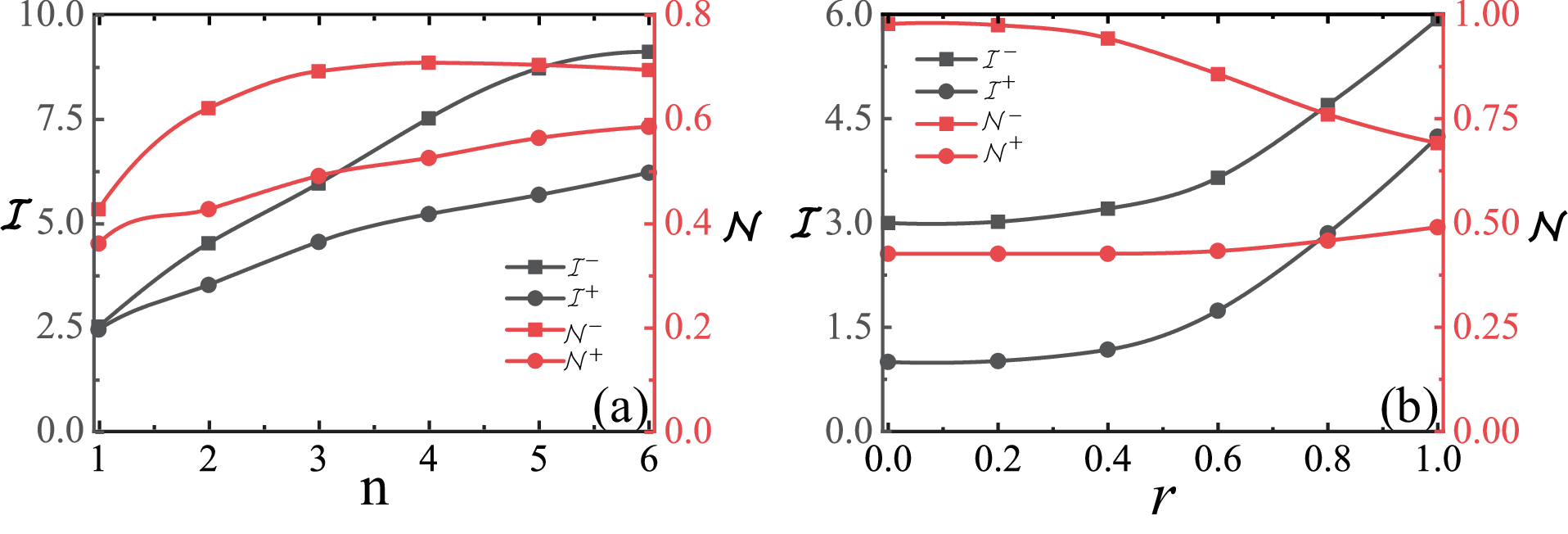}}}
\caption{The dependences of the macroscopicity $\mathcal I$ and Wigner negativity  $\mathcal N$ of the non-Gaussian mechanical states on the number $n$ (a) and the squeezing $r$ (b) in the MPS ($\mathcal I^-$, $\mathcal N^-$) and MPA ($\mathcal I^+$, $\mathcal N^+$) operations, with $r=1$ in (a) and $n=3$ in (b).}
\label{fig9}
\end{figure}

\begin{figure}[H]
\centerline{\hspace{0cm}\scalebox{0.27}{\includegraphics{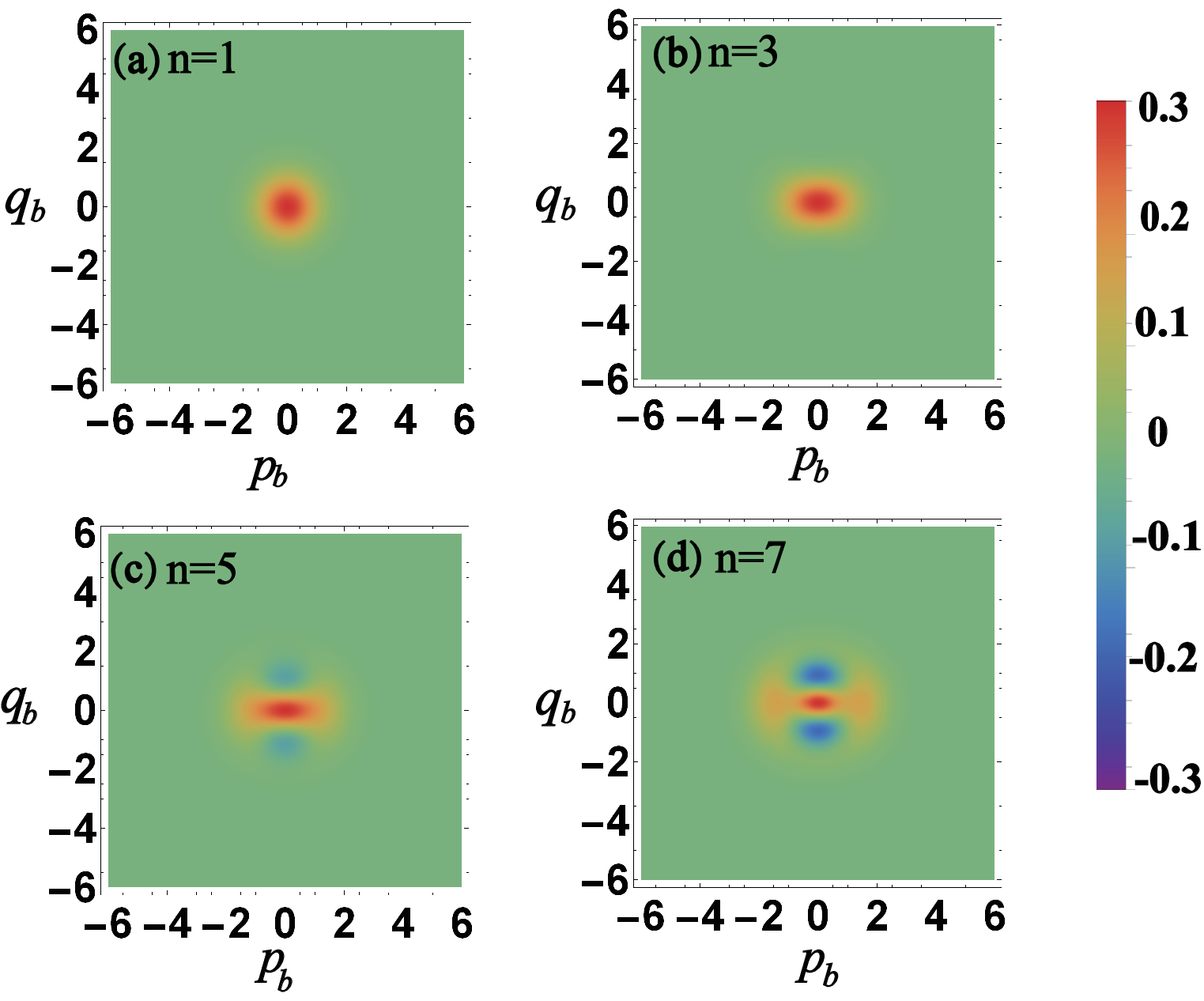}}}
\vspace{-0.2cm}
\caption{The same as Fig.7 but for catalyzing $n$ photons, with the squeezing parameter $r=0.1$.}
\label{fig10}
\end{figure}

In Figs.7 and 8, the density plots of the Wigner functions $W_b(q_b,p_b)$ of the density operators $\hat\rho_b$ are presented for adding and subtracting $n$ photons, respectively, with the homodyne detection outcome $q_{\tilde{a}}=0$. We see that the Wigner functions exhibit strong negativity and obvious interference patterns. This shows that
mechanical non-Gaussian states with strong nonclassicality are generated remotely.
Obviously, the approximate odd (even) cat states can be generated for the mechanical oscillator when the system driven by the blue-detuned laser and the photon number $n$ is odd (even). This is due to the non-Gaussian steerable correlation, induced by the multiphoton operations on the OM two-mode squeezed vacuum state, that enables homodyne projecting the mechanical oscillator into the cat states. Since the OM two-mode squeezed vacuum state is a superposition only of states in which the two modes contain the same number of photons and phonons, the odd (even) photon subtraction or addition leads to a change in  photon and phonon numbers contained in the superpositioned states (de-Gaussian process). Then, the homodyne detection on the optical output field, with the detection result $q_{\tilde{a}}=0$, ``filters" out the odd (even) phononic number states from the superposition of the conditional mechanical state $\hat \rho_b(q_{\tilde{a}})$ , resulting in the odd (even ) cat state of the mechanics, as the Hermite polynomial of odd order is equal to zero for $q_{\tilde a}=0$ [i.e., the Hermite polynomial $H_{k}(q_{\tilde{a}}=0)=0$ for odd number $k$ in Eqs.(\ref{rms1}) and (\ref{rms2})]. Note that the above de-Gaussian is necessary; otherwise the homodyne detection merely results in Gaussian mechanical conditional states.
For instance, we have the fidelity $F_-^{\rm max}\approx 0.98$, with respect to an ideal even cat state with the amplitude $\alpha\approx3.8$, for subtracting $n=6$ photons. In this case, the conditional state $|\psi_{b}\rangle\approx0.13|6\rangle+0.28|8\rangle+0.38|10\rangle+0.43|12\rangle$, which is very close to the ideal even cat state $|\psi_{\alpha=3.8}\rangle{\rm cat}\approx0.12|6\rangle+0.23|8\rangle+0.34|10\rangle+0.43|12\rangle$. Therefore, large-size cat states can be remotely achieved with the present scheme.
However, in the continuous-drive case, the steady-state OM entangled state is merely a highly mixed state (without the photon-phonon pairs contained in the two-mode squeezed vacuum state as in the pulse-drive case), and thus the combination of the photon subtraction (addition) operation and the homodyne detection fails to make the superposition of odd or even phononic number states. Therefore, the scheme with pulsed drive is more favorable for the remote generation of large-size mechanical cat states via the multiphoton operation and homodyne detection in cavity OM systems. For example, in the continuous-drive case, from Fig.5(a), with the photon subtraction $n=3$ and $\kappa_c=4\omega_m$, the steering $S_{FI}^{\tilde{a}\rightarrow b}=0.394$. The same amount of steering is also achieved in the pulse case with $r\approx 0.175$, for the photon subtraction $n=3$. But we have the Wigner negativity $\mathcal N\approx 0.97$ of the conditional mechanical states in the pulse case, which is much bigger than that in the continuous-drive case. This shows that even with the same non-Gaussian OM steering, the pulse scheme is more efficient for generating negative-Wigner mechanical states than the continuous-drive scheme.

Further, the size and negativity of the generated cat states increase as $n$ increases.
Our results show that multiphoton addition or subtraction can enhance the size of the remote cat states considerably. This is because that the photon projection measurement with larger photon number and subsequent homodyne detection on the mode ${\tilde a}$ can lead the maximal probability distribution in the Fock space of the mechanical mode $b$ to shift towards larger Fock states.
Compared Fig.\ref{fig7} with Fig.\ref{fig8}, the larger macroscopic quantum superpositions can be created by the MPS than the MPA for the given squeezing $r$ and photon number $n$. This can be judged with the macroscopicity of quantum superpositions, defined as \cite{I}
\begin{align}
\mathcal I=\frac{\pi}{2}\int W_b(\beta)(-\frac{\partial^{2}}{\partial\beta\partial\beta^*}-1)
W_b(\beta)d^{2}\beta.
\end{align}
As shown in Fig.9 which plots the macroscopicity $\mathcal I$ of the non-Gaussian mechanical states and negativity $\mathcal N$ of the corresponding Wigner functions, the macroscopicity increases with the photon number $n$ and squeezing $r$, and we have $\mathcal I^->\mathcal I^+$, i.e., the macroscopicity ($\mathcal I^-$) of the mechanical superpositions induced by the MPS is larger than that ($\mathcal I^+$) by the MPA. This can be understood from two aspects: first, different from photon addition, the subtraction operation directly ``filters" the small Fock states (smaller than $n$) from the superpositions of the state $|\psi\rangle_{\rm {\tilde a}b}$, resulting in the superpositions involving larger Fock states and increasing macroscopicity; second, this is due to the light-to-mechanics steering by the MPS is larger than that by the MPA, as shown in Fig.\ref{fig3}, which allows for the generation of larger cat states via the MPS. In addition, the Wigner negativity $\mathcal N$ increases to saturation with the increasing of the number $n$ of added and subtracted photons, as shown in Fig.9 (a), but the negativity for $n=3$ in the MPS decreases with the squeezing $r$ in Fig.9 (b). This is because that for very small of the squeezing, the mechanical oscillator is prepared by the MPS operations and homodyne detection in an approximate three-phonon Fock state, which possesses larger negativity. It is also shown that the negativity $\mathcal N^-$ of the nonclassical non-Gaussian mechanical states induced by the MPS is larger than  $\mathcal N^+$ by the MPA, similar to the property of the macroscopicity.

\begin{figure}
\centerline{\hspace{0cm}\scalebox{0.22}{\includegraphics{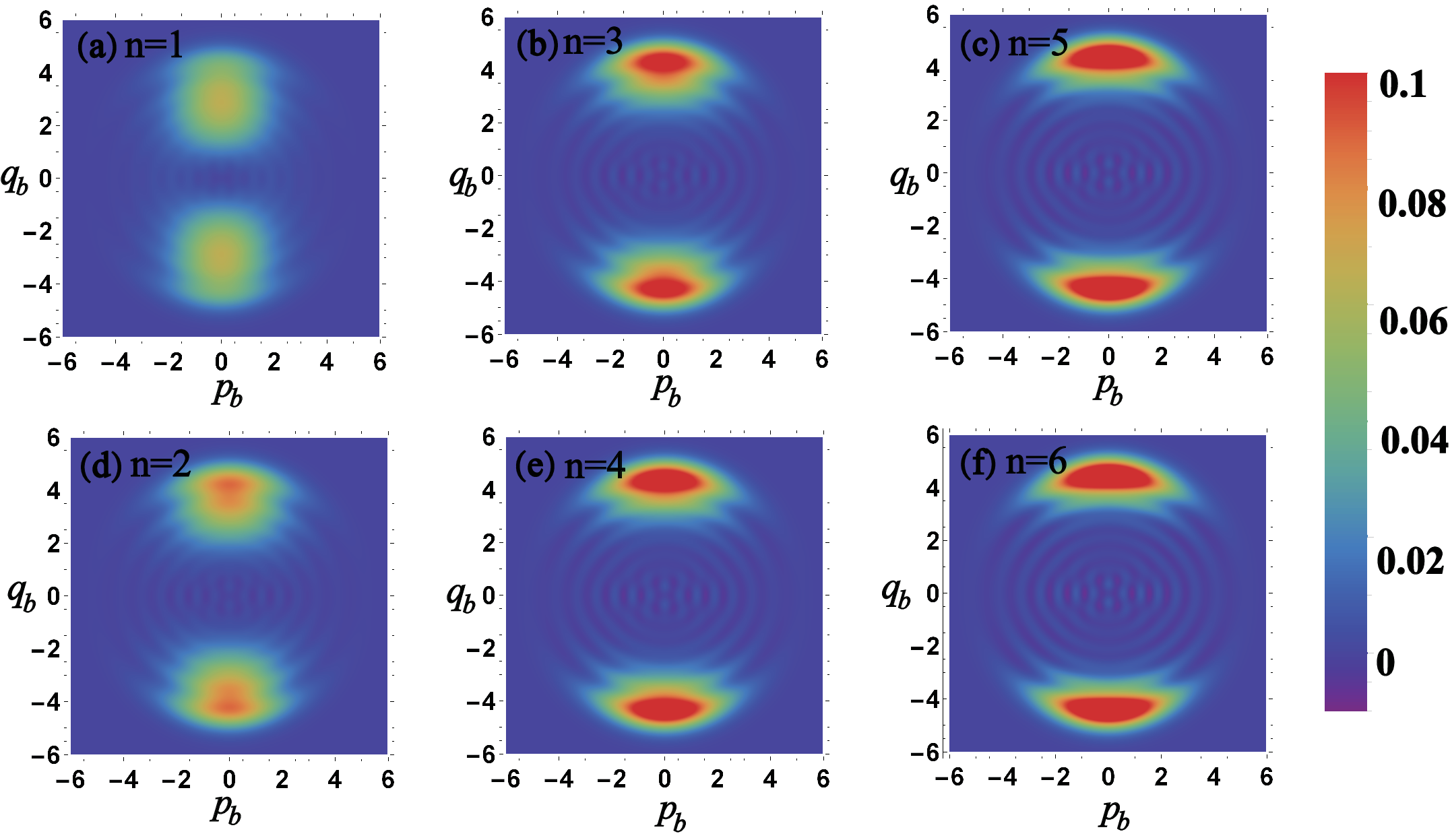}}}
\caption{The density plots of the Wigner function $W_b(q_b,p_b)$ of the mechanical state $\hat \rho_b(q_{\tilde a}=0)$ in the continuous-drive case under the MPS operation, with  $\kappa_c
=4\omega_m$ and other parameters are the same as in Fig.\ref{fig5}.}
\label{fig11}
\end{figure}

\begin{figure}
\centerline{\hspace{0cm}\scalebox{0.22}{\includegraphics{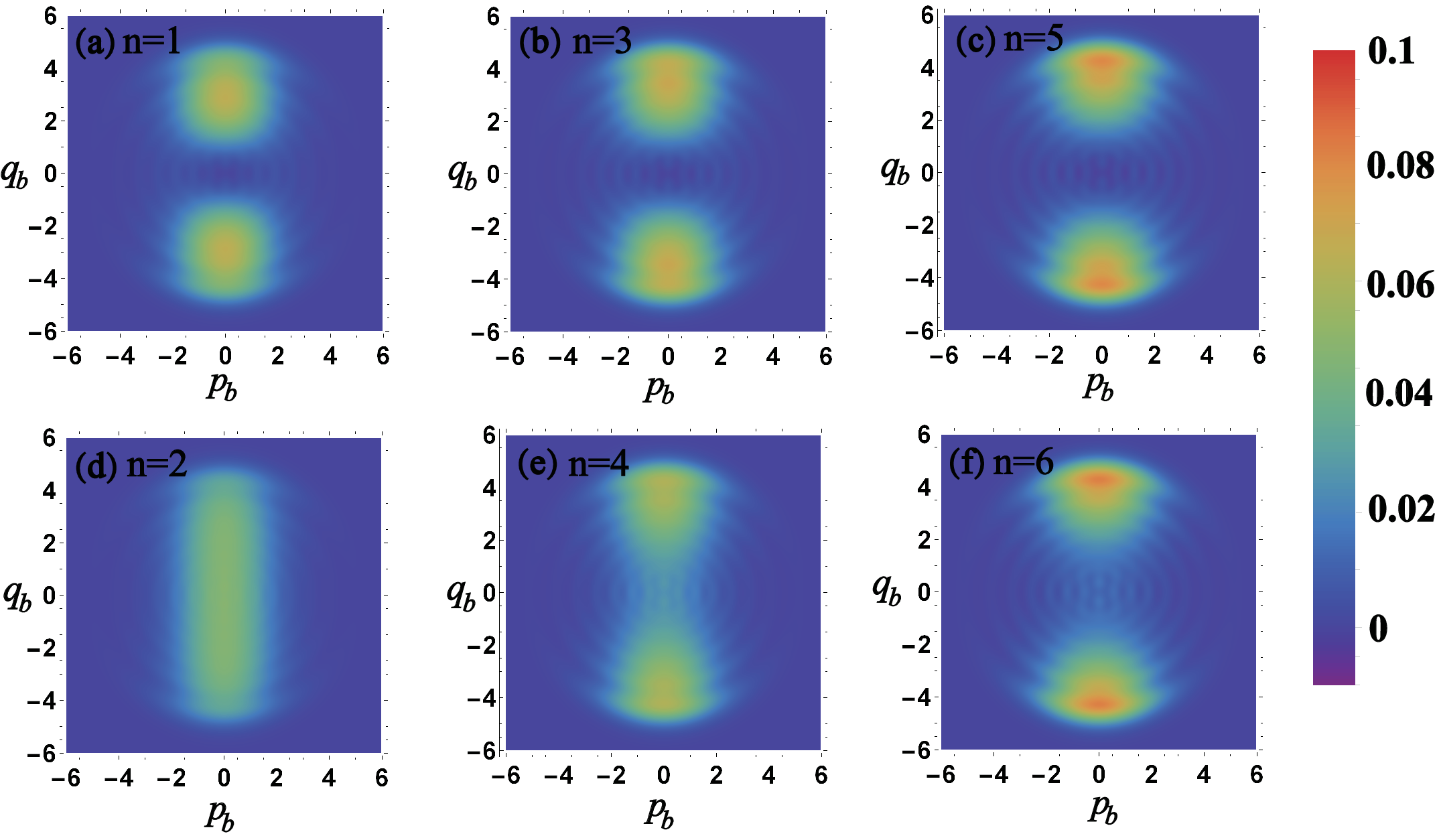}}}
\caption{The same as Fig.\ref{fig11} but for adding $n$ photons.}
\label{fig12}
\end{figure}

\begin{figure}
\centerline{\hspace{0cm}\scalebox{0.27}{\includegraphics{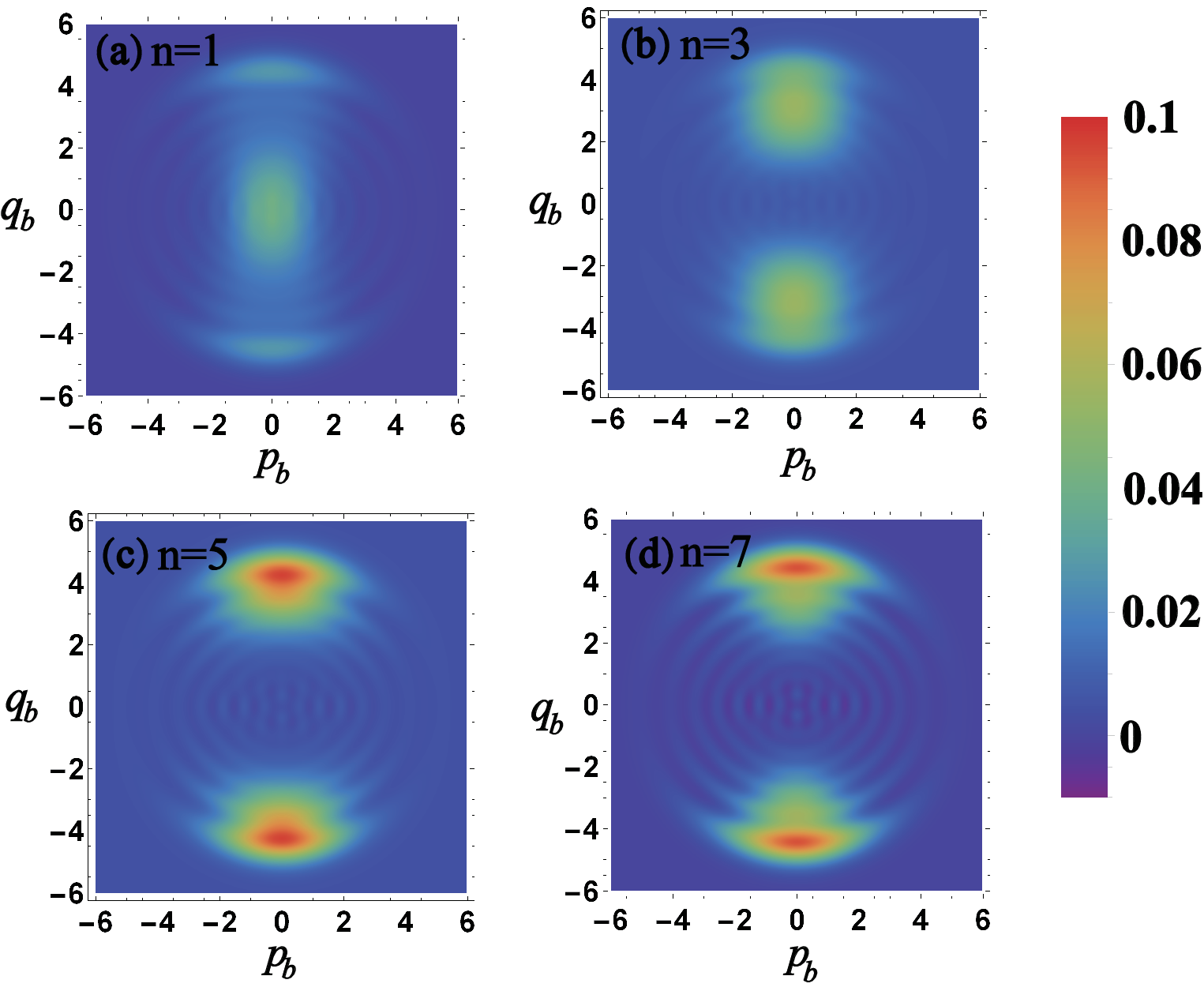}}}
\caption{The same as Fig.\ref{fig11} but for catalyzing $n$ photons.}
\label{fig13}
\end{figure}

In Fig. \ref{fig10}, the density plots of the Wigner functions with the MPC operations are presented for the case of pulsed drive. It is shown that only mechanical kitten states can be achieved, with the transmissivity $T=0.1$ and squeezing $r=0.1$. The amplitude of the generated cat states is very small even for $n=7$, as shown in the fig.10(d), compared to those in Fig.\ref{fig7} and Fig.\ref{fig8}, since the non-Gaussian OM steering induced by the MPC operation is much smaller than those by the MPS and MPA operations.

We finally study the Wigner functions of the mechanical states in the case of the continuous drive. As illustrated in Figs.\ref{fig11}-\ref{fig13}, which respectively correspond to the situations of the MPS, MPA and MPC operations, the Wigner functions also exhibit negativity, indicating the nonclassicality non-Gaussian mechanical states can still be achieved via the homodyne detection. However, since the achievable non-Gaussian OM steering with the continuous drive is much weaker than that with the pulsed drive, weaker Wigner negativity in the mechanical states is thus resulted. This further suggests that strong non-Gaussian quantum steerable correlations are necessary for the remote generation of the non-Gaussian states with significant nonclasscality. Moreover, unlike the pulse-drive case in which an OM two-mode squeezed vacuum state is prepared, the steady-state OM entangled state via continuous driving is a highly mixed state (not in a superposition of states in which the mechanical and optical modes contain the same number of photons and phonons),  and thus the combination of the photon subtraction (addition) and the homodyne detection fails to make the superposition of odd or even phononic number states. Consequently, as we show, the pulsed drive is more favorable for the remote generation of the large-size mechanical cat states in the cavity OM systems.

\section{Discussion and Conclusion}
Before concluding, let us briefly discuss the feasible experimental realizations of the proposed scheme, based on currently MPS, MPA, and MPC operations available in optical systems. Experimentally, optomechanical parametric downconversion and photon-phonon entanglement with pulsed drive in photonic-crystal nano-optomechancial systems have been achieved \cite{exper1}. In the experiments, the mechanical frequency $\omega_m/2\pi\sim5$ GHz and linewidth $\gamma_m/2\pi\sim300$ kHz, giving the quality factor $Q\sim2\times 10^4$. With the cryostat at a temperature of 60 mK, the mean thermal excitation number $\bar n_{th}<1$.
By adjusting appropriate pulse amplitude and duration, the parameter requirements for achieving light-mechanical two-mode squeezed states can be satisfied. For the case of continuous drive, low frequency mechanical oscillator may be chosen to operate in the sideband-unresolved regime, as in the experiments\cite{exper2,exper3} in which the mechanical oscillator is cooled to a state with $\bar n_{th}\sim 5$ via feedback cooling.

In conclusion, we consider the generation of the non-Gaussian OM steering via non-Gaussian multphoton operations in cavity optomechanical systems and the remote preparation of large-size motional Schr\"{o}dinger cat states with the help of the non-Gaussian quantum steerable correlations. We consider that three typical kinds of multiphoton operations, i.e., MPS, MPA, and MPC, are employed on the output field from the OM cavity which is assumed to be driven by pulse or continuous lasers. We find that these multiphoton operations can lead to non-Gaussian OM quantum steerable correlations, and the steering is significantly enhanced with an increasing number $n$ of photons in the multiphoton operations. We reveal that the FI-based steering criterion is much more effective for detecting the non-Gaussian steering than the well-known Reid's criterion.
We further show that the strong OM steering enables the remote preparation of large-size Schr\"{o}dinger odd or even cat states of the mechanical oscillator by homodyne detection. The amplitudes of the cat states also increase significantly with the photon number $n$ in the multiphoton operations. Our results reveal the properties of non-Gaussian steering generated by multiphoton operations, and strong non-Gaussian steering is requisite for distantly achieving large-size cat states by homodyne detection. The OM non-Gaussian steering and the quantum superpositions of macroscopic mechanical resonators hold promises for fundamental tests in quantum mechanics and practical applications in quantum science.

\section*{ACKNOWLEDGMENT}
This work is supported by the National Natural Science Foundation of China (No. 12174140) and the Fundamental Research Funds for the Central Universities (No. CCNU24JC014).

\setcounter{equation}{0}
\renewcommand\theequation{A\arabic{equation}}

\section*{APPENDIX: The derivation of Eq.(\ref{eq3})}
When the virtual cavity field $\hat a_{\mu}^{out}$ (filtered mode from the continuous output $\hat a^{out}$) is cascadely coupled to the OM cavity with the time-dependent coupling $g_{\mu}(t)$, the Heisenberg-Langevin equation for the virtual cavity field $\hat a_\mu^{out}$ is given by
\begin{align}
\dot{\hat a}_{\mu}^{out}=\frac{|g_{\mu}(t)|^{2}}{2}\hat a_{\mu}^{out}-g_{\mu}(t)\hat a^{out}(t).
\end{align}
It formal solution can be derived as
\begin{align}
\hat a_{\mu}^{out}(t)&=\hat a_{\mu}^{out}(0)e^{-\frac{1}{2}\int_0^t dt'|g_{\mu}(t')|^{2}}\nonumber\\
&-\int_0^tdt' g_{\mu}(t')\hat a^{out}(t')e^{-\frac{1}{2}\int_{t'}^{t} dt''|g_{\mu}(t'')|^{2}},
\end{align}
which can lead to
\begin{align}
\hat a_{\mu}^{out}=\int_0^{\infty} \mu^*(t')\hat a^{out}(t')dt',
\end{align}
with the temporal mode function $\mu(t)$ defined as
\begin{align}
\mu^*(t)=-g_{\mu}(t)e^{-\frac{1}{2}\int_t^{\infty} dt''|g_{\mu}(t'')|^{2}}.
\label{a4}
\end{align}
The above equation gives
\begin{align}
\int_0^t dt'|g_{\mu}(t')|^{2} e^{-\int_{t'}^{\infty} dt''|g_{\mu}(t'')|^{2}}=\int_0^t dt'|\mu(t')|^{2},
\end{align}
and
\begin{align}
e^{-\int_t^{\infty} dt'|g_{\mu}(t')|^{2}}=\int_0^t dt'|\mu(t')|^{2}.
\label{a6}
\end{align}
Substituting (\ref{a6}) in (\ref{a4}), we have
\begin{align}
g_{\mu}(t)=-\frac{\mu^*(t)}{\sqrt{\int_0^{t} dt'|\mu(t')|^2}}.
\end{align}

\end{document}